\documentclass[12pt,amsmath,amssymb,aps]{revtex4-1}
\usepackage{graphicx,epsfig,dcolumn,bm,epic,eepic,float}
\usepackage{amsmath}
\usepackage[T1]{fontenc}
\usepackage[utf8]{inputenc}
\usepackage{mathtools} 
\usepackage{pifont}
\usepackage{amsmath}
\usepackage{latexsym}
\usepackage{color}
\usepackage{cancel}
\usepackage{scrextend}
\usepackage{amsmath}
\usepackage{amsthm}
\usepackage{eucal}
\usepackage{amssymb}
\usepackage{mathrsfs}
\usepackage[bottom]{footmisc}
\usepackage{makeidx,shortvrb,latexsym}
\usepackage{epstopdf}
\usepackage{mathtools}
\usepackage{ragged2e}
\usepackage[T1]{fontenc}
\newcommand {\be}{\begin{equation}}
\newcommand {\ee}{\end{equation}}
\newcommand {\ba}{\begin{eqnarray}}
\newcommand {\ea}{\end{eqnarray}}
\newcommand {\bea}{\begin{eqnarray}}
\newcommand {\eea}{\end{eqnarray}}

\pdfoutput=1

\numberwithin{equation}{section}
\begin{document}

\begin{flushright}
MAN/HEP/2019/02\\
April 2019
\end{flushright}
\vspace*{0.7cm}

\title{{\LARGE Quartic Coupling Unification} \\[3mm]
{\LARGE in the Maximally Symmetric 2HDM}\\[1mm]
${}$}

\author{\large Neda Darvishi$\,$} \email{neda.darvishi@manchester.ac.uk}
\author{\large Apostolos Pilaftsis$\,$}\email{apostolos.pilaftsis@manchester.ac.uk}

\affiliation{~~~~~~~~~~~~~~~~~~~~~~~~~~~~~~~~~~~~~~~~~~~~~~~~~~~~~~~~~~~~${}$\vspace{-3mm}\\
Consortium for Fundamental Physics, School of Physics and
  Astronomy,\\University of Manchester, Manchester M13 9PL, United
  Kingdom}

\begin{abstract}
${}$

\centerline{\bf ABSTRACT} \medskip

\noindent
We consider the Maximally Symmetric Two-Higgs Doublet Model (MS-2HDM)
in which the so-called Standard Model (SM) alignment can be naturally
realised as a consequence of an accidental SO(5) symmetry in the Higgs
sector.  This symmetry\- is broken (i)~explicitly by
renormalization-group~(RG) effects and (ii)~softly by the bilinear
scalar mass term~$m^2_{12}$.  We find that in the MS-2HDM all quartic
couplings can unify at large RG scales
$\mu_X \sim 10^{11}\,$--$\,10^{20}$\,GeV.  In particular, we show that
quartic coupling unification can take place in two different
conformally invariant points, where all quartic couplings vanish.  We
perform a vacuum stability analysis of the model in order to ensure
that the electro\-weak vacuum is sufficiently long-lived.  The MS-2HDM
is a minimal and very predictive extension of the SM governed by only
three additional parameters: the unification scale~$\mu_X$,
the charged Higgs mass $M_{h^{\pm}}$ (or $m^2_{12}$) and $\tan\beta$,
which allow one to determine the entire Higgs sector of the model.
In terms of these input parameters, we present illustrative
predictions of misalignment for the SM-like Higgs-boson couplings
to the $W^\pm$ and $Z$ bosons and, for the first time, to the top and bottom quarks. 

\end{abstract}

\maketitle

\section{Introduction}
\label{sec:intro} 

Despite intense scrutiny, the Standard Model (SM) has proven to be
very successful in describing the fundamental interactions of Particle
Physics~\cite{Glashow:1961tr,Weinberg:1967tq,Salam1968}.  The
discovery of the Higgs particle~\cite{Englert:1964et,Higgs:1964pj} 
at the CERN Large Hadron Collider
(LHC) was one of the most important achievements towards a minimal
ultra-violet (UV) completion of the SM
\cite{Aad:2012tfa,Chatrchyan:2012xdj}. In spite of its great success,
the search for new physics beyond the SM still has strong theoretical
and experimental motivations, and opens up new possibilities including
the study of non-standard scenarios with extended Higgs sectors.
However, latest LHC data dictate that the observed Higgs boson must
interact with the electroweak (EW) gauge bosons ($Z, \,W^\pm$) with
coupling strengths that are very close to their SM values
\cite{ATLAS1,CMS1}. This simple fact puts severe limits on possible
scalar-sector extensions of the SM.

One class of minimal extensions of the SM is the Two-Higgs Doublet
Model (2HDM), where the SM scalar sector is extended by a second Higgs
doublet \cite{Lee1,Branco1}. This model can, in principle, account for
a SM-like Higgs boson, and contains additionally one charged and two
neutral scalars whose observation could be within reach of the
LHC~\cite{Arbey:2017gmh,Hanson:2018uhf}.

In the 2HDM, one may have four different types of Yukawa interactions
mediating no Flavor Changing Neutral Currents (FCNCs) at the tree
level  \cite{Glashow:1976nt,Hall,Barger:1989fj,Pich}. More explicitly, in Type-I
(Inert-type), all the fermions couple to the first doublet~$\Phi_1$
and none to the second doublet~$\Phi_2$. In Type-II (MSSM-Type), the
down-type quarks and the charged leptons couple to~$\Phi_1$ and the
up-type quarks couple to~$\Phi_2$. In Type-III (flipped-model), the
down-type quarks couple to $\Phi_1$ and the up-type quarks and the
charged leptons couple to $\Phi_2$. In Type-IV
(lepton-specific-model), all quarks couple to $\Phi_1$ and the charged
leptons couple to $\Phi_2$.  As mentioned above, in all these
different settings of the 2HDM, the couplings of the SM-like Higgs boson to the
electroweak gauge bosons ($Z,\,W^\pm$) must be very close to those
predicted by the SM, so as to be in agreement with the current Higgs
signals at the LHC. This is only possible within the so-called SM
alignment limit of the 2HDM
\cite{Ginzburg:1999fb,Chankowski:2000an,Delgado:2013zfa,Carena:2013ooa,Dev:2014yca,Bernon:2015qea,Benakli:2018vqz,Lane:2018ycs}. 
In particular, in the Type-II 2HDM, the
couplings of the SM-like Higgs boson to vector bosons lie within
10$\%$ of the SM value at 95$\%$ CL
\cite{Grinstein:2013npa,Baglio:2014nea,Darvishi:2017bhf}.

In this paper, we consider the simplest realisation of a Type-II 2HDM,
the so-called Maximally Symmetric Two-Higgs Doublet Model (MS-2HDM).
In this model, the aforementioned SM alignment can emerge naturally as
a consequence of an accidental SO(5) symmetry in the Higgs sector~
\cite{Pilaftsis:2011ed,Dev:2014yca,Pilaftsis:2016erj,Dev:2017org},
without resorting to {\em ad-hoc} arrangements among the parameters
of the theory ~\cite{Georgi:1978ri,Gunion:2002zf,CP2005,Carena:2013ooa,Haber:2015pua,Grzadkowski:2018ohf}. The SO(5) symmetry is broken explicitly by
two sources: (i)~by renormalization-group~(RG) effects and (ii)~softly
by the bilinear scalar mass term~$m^2_{12}$. A~remarkable feature of
the MS-2HDM is that all quartic couplings can unify at very large
scales $\mu_X \sim 10^{11}\,$--$\,10^{20}$\,GeV, for a wide range of
$\tan\beta$ values and charged Higgs-boson masses.  In particular, we
find that quartic coupling unification can happen in two different
conformally invariant points, where all quartic couplings vanish. The
first conformal point is at relatively low-scale typically of
order~$10^{11}$~GeV, while the second one is at high scale close to
the Planck scale~$\sim 10^{19}$~GeV. Most remarkably, we find that the
MS-2HDM becomes a very predictive extension of the SM which is
governed by only three additional parameters: the quartic coupling unification
scale~$\mu_X$, the charged Higgs mass $M_{h^{\pm}}$ (or $m^2_{12}$)
and $\tan\beta$. These three parameters also suffice to determine the entire
Higgs-mass spectrum of the model.  By means of these  input
parameters, we are able to obtain definite predictions of misalignment for the
SM-like Higgs-boson couplings to the $W^\pm$ and $Z$ bosons and to the
top- and bottom-quarks, which might be testable at future precision
high-energy colliders.

The layout of the paper is as follows. After this introductory
section, Section \ref{alignment} briefly reviews the basic features of
the 2HDM and discusses the conditions for achieving exact SM
alignment. In Section \ref{maxsym}, we describe the MS-2HDM, thereby
illuminating the origin of {\em natural} SM alignment. We also outline
the breaking pattern of the SO(5) symmetry, which results from the
soft-breaking mass $m^2_{12}$ and the RG effects, and discuss its
implications for the Higgs-mass spectrum. In Section \ref{QCU}, we
analyse in more detail the impact of RG effects up to two loops on all
relevant quartic couplings by considering their running from the
quartic coupling unification scale $\mu_X$ to the charged Higgs-boson
mass.  In particular, we show that the running quartic couplings can
be unified at two different conformally-invariant points. In the same
section, we present illustrative predictions for the unification of
all quartic couplings for typical values of $\tan\beta$ and charged
Higgs-boson masses. Section~\ref{misa} presents our misalignment
predictions for Higgs-boson couplings to gauge bosons and top- and bottom-quarks. In
Section~\ref{vac}, we analyse the EW vacuum lifetime of
the MS-2HDM by considering the bounce solution to the Euclidean
equation of motion for the classical potential, with a negative running quartic coupling $\lambda_2$. We estimate the EW vacuum
lifetime~$\tau$ which turns out to be adequately long, being many orders
of magnitude larger than the age of the Universe (in the absence of
Plank-scale suppressed operators \cite{Branchina:2013jra,Branchina:2018xdh}). Finally,
Section \ref{con} contains our conclusions.

\section{Type-II 2HDM and SM Alignment} \label{alignment}

The Higgs sector of the 2HDM is described by two scalar SU(2) 
doublets,
\begin{equation}
{\Phi}_1\ =\ \left(
 \begin{matrix}
\phi_1^+\\
 \phi_1^0
 \end{matrix}\right)\;,\qquad 
{\Phi}_2\ =\ \left(
 \begin{matrix}
\phi_2^+\\
 \phi_2^0
 \end{matrix}\right)\;.
\end{equation} 
In terms of these doublets, the most
general SU(2)$_L\otimes$U(1)$_Y$-invariant Higgs potential is given by
\begin{eqnarray}
  \label{eq:V2HDM}
V &=& - \mu_1^2 ( \Phi_1^{\dagger} \Phi_1) - \mu_2^2 ( \Phi_2^{\dagger} \Phi_2) 
 - \Big[ m_{12}^2 ( \Phi_1^{\dagger} \Phi_2)\: +\: {\rm H.c.}\Big] \nonumber \\
 &+& \lambda_1 ( \Phi_1^{\dagger} \Phi_1)^2 + \lambda_2 ( \Phi_2^{\dagger} \Phi_2)^2
 + \lambda_3 ( \Phi_1^{\dagger} \Phi_1)( \Phi_2^{\dagger} \Phi_2)
 + \lambda_4 ( \Phi_1^{\dagger} \Phi_2)( \Phi_2^{\dagger} \Phi_1) \nonumber \\
 &+& \bigg[\, {1 \over 2} \lambda_5 ( \Phi_1^{\dagger} \Phi_2)^2
 + \lambda_6 ( \Phi_1^{\dagger} \Phi_1)( \Phi_1^{\dagger} \Phi_2)
 + \lambda_7 ( \Phi_1^{\dagger} \Phi_2)( \Phi_2^{\dagger} \Phi_2)\: +\:
     {\rm H.c.} \bigg]\;,
\end{eqnarray}
where the mass term $m^2_{12}$ and quartic couplings $\lambda_5$,
$\lambda_6$ and $\lambda_7$ are complex parameters.  Instead, the
remaining mass terms, $\mu^2_{1}$ and $\mu_{2}^2$, and the quartic
couplings $\lambda_{1,2,3,4}$ are real. Of these 14 theoretical
parameters, only 11 are physical, since three can be removed away
using an SU(2) reparameterisation of the Higgs doublets $\Phi_1$ and
$\Phi_2$~\cite{CP2005}.
 
In the present article, we will restrict our attention to CP
conservation and to CP-conserving vacua.  In the Type-II 2HDM, both
scalar doublets $\Phi_1$ and $\Phi_2$ receive nonzero vacuum
expectation values (VEVs). Specifically, we have
$\langle \phi_1^0\rangle= v_1/\sqrt{2}$ and
$\langle \phi_2^0 \rangle = v_2/\sqrt{2}$, where $v_{1,2}$ are
non-zero and $v \equiv \sqrt{v_1^2 + v_2^2} $ is the VEV of
the SM Higgs doublet. The minimization conditions resulting from the 2HDM
potential in~\eqref{eq:V2HDM} give rise to the following relations:
\begin{eqnarray}
\mu_{1}^2\ &=&\ m_{12}^2 t_\beta - {1\over 2} v^2 {c^{2}_\beta} (2
            \lambda_1+3 \lambda_6 {t_\beta} + \lambda
            _{345}{t^2_\beta}+\lambda_7 {t^3_\beta } )\,,\\ 
\mu_{2}^2\ &=&\ m_{12}^2 {t^{-1}_\beta} - {1\over 2} v^2 {s^{2}_\beta} (2
            \lambda_2+3 \lambda_6 {t^{-1}_\beta} + \lambda
            _{345}{t^{-2}_\beta}+\lambda_7 {t^{-3}_\beta } )\,, 
\end{eqnarray}
where $s_{\beta} \equiv \sin{\beta}$, $c_{\beta} \equiv \cos{\beta}$,
$t_\beta\equiv\tan\beta =v_2/v_1$ and
${\lambda }_{345}\equiv {\lambda }_3+{\lambda }_4+{\lambda }_5$.
Following the standard linear expansion of 
the two scalar doublets $\Phi_j$ (with $j =1,2$)
about their VEVs, we may conveniently re-express them as
\begin{equation}
\Phi_j \  =\ \left(
 \begin{matrix}
 \phi_j^+ \\
 {1 \over \sqrt{2}} (v_j + \phi_j + i\phi_j ^0)
 \end{matrix}
 \right)\;.
\end{equation}
After spontaneous symmetry breaking (SSB), the standard EW gauge fields,
the $W^{\pm}$ and $Z$ bosons, acquire their masses from the three
would-be Goldstone bosons $(G^{\pm},G^0)$~\cite{Goldstone}. As a
consequence, the model has only five physical scalar states: two
CP-even scalars ($h$,$H$), one CP-odd scalar $(a)$ and two charged
bosons ($h^{\pm}$). The mixings in the CP-odd and charged scalar sectors
are individually governed by the same angle $\beta$,
\begin{eqnarray}
 G^{\pm} &=& c_\beta \phi_1^{\pm} + s_\beta \phi_2^{\pm}\;, \,
             \,\,\,\,\, \,\,\,\,\, \,\,\,\, h^{\pm} = -s_\beta
             \phi_1^{\pm} + c_\beta \phi_2^{\pm}\;, \\ \nonumber 
 G^0 &=& c_\beta \phi_1 + s_\beta \phi_2\;, \,\,\,\,\,\,\,
         \,\,\,\,\,\,\, \,\,\,\,\,\,\, a = -s_\beta \phi_1 + c_\beta
         \phi_2\;. 
\end{eqnarray}
Correspondingly, the masses of the $h^\pm$ and $a$ scalars are given by
\begin{eqnarray}
M_{h^{\pm}}^2 \ &=&\ {m_{12}^2 \over s_{\beta} c_{\beta}} - {v^2 \over 2} (\lambda_4 + \lambda_5)
 + {v^2 \over 2 s_{\beta} c_{\beta}} (\lambda_6 c_{\beta}^2 
 + \lambda_7 s_{\beta}^2)\;,
\nonumber \\
M_a^2 \ &=&\ M_{h^{\pm}}^2 + {v^2 \over 2} (\lambda_4 - \lambda_5)\;.
\end{eqnarray}

To obtain the masses of the two CP-even scalars, $h$ and $H$, we need
to diagonalise the two-by-two CP-even mass matrix~$M^2_S$,
\begin{equation}
M_S^2=\left(
\begin{matrix}
A& C\\
 C& B
 \end{matrix}\right)\; ,
 \label{abc}
\end{equation}
which may explicitly be written down as
 \begin{eqnarray}
M_S^2\ =\ M_a^2 \left(
 \begin{matrix}
 s_{\beta}^2 & -s_{\beta}c_{\beta} \\
 -s_{\beta}c_{\beta} & c_{\beta}^2
 \end{matrix}
 \right) 
 + v^2 \left(
 \begin{matrix}
 2 \lambda_1 c_{\beta}^2 + \lambda_5 s_{\beta}^2 + 2 \lambda_6 s_{\beta} c_{\beta} &
 \lambda_{34} s_{\beta} c_{\beta} + \lambda_6 c_{\beta}^2 + \lambda_7 s_{\beta}^2 \\
 \lambda_{34} s_{\beta} c_{\beta} + \lambda_6 c_{\beta}^2 + \lambda_7 s_{\beta}^2 &
 2 \lambda_2 s_{\beta}^2 + \lambda_5 c_{\beta}^2 + 2 \lambda_7 s_{\beta} c_{\beta}
 \end{matrix}
 \right),\nonumber 
\end{eqnarray}
with $\lambda_{34} \equiv \lambda_{3} + \lambda_{4}$. The mixing angle
$\alpha$ necessary for the diagonalisation of $M_S^2$ may be determined by 
\begin{equation}
   \label{alpha}
\tan 2\alpha\ =\ {2C \over A-B}\; .
\end{equation}

The SM Higgs field may now be identified by the linear field combination,
\begin{equation}
   \label{g-c}
H_{\text{SM}}\ =\ \phi_1 \cos \beta + \phi_2 \sin \beta\ 
 =\ H \cos (\beta - \alpha) + h \sin (\beta - \alpha)\; .
\end{equation}
In this way, one can obtain the SM-normalised couplings of the CP-even $h$
and $H$ scalars to the EW gauge bosons ($V = W^{\pm}, Z$) as follows:
\begin{equation}
g_{hVV} = \sin (\beta - \alpha)\;, \qquad g_{HVV} = \cos (\beta - \alpha)\;,
\end{equation}
with $g_{H_{\rm SM}VV} = 1$ by definition.  In similar manner, we
may derive the SM-normalised couplings of the
CP-even and CP-odd scalars to up-type and down-type quarks. These
couplings are exhibited in Table~\ref{tab}. 

\begin{table*}[t]
 \centering
 \small
 \begin{tabular}{l c c c c c c c c c c }
 \hline\hline
$S$ &&& $g_{SVV}$ ($V=W^{\pm},Z$) &&& $g_{Suu}$ &&&& $g_{Sdd}$ \\
 \hline
 $h$ &&& $\sin(\beta-\alpha)$ &&& $ \sin(\beta-\alpha)+\frac{1}{\tan\beta}\cos(\beta-\alpha)$ &&&& $\sin(\beta-\alpha)-\tan\beta\cos(\beta-\alpha)$ \\
 $H$ &&& $\cos(\beta-\alpha)$&&&$ \cos(\beta-\alpha)-\frac{1}{\tan\beta}\sin(\beta-\alpha)$ &&&& $\cos(\beta-\alpha)+\tan\beta\sin(\beta-\alpha)$\\
 $a$ &&& $ 0$ &&& $- i\gamma_5 \cot\beta$ &&&& $- i\gamma_5 \tan\beta $\\
 \hline\hline
 \end{tabular}
 \caption{\it
 Tree-level couplings of a neutral scalar boson $S$ (with $S = h,\,
 H,\, a$) to the $W^\pm$ and $Z$ bosons
 and to quarks in the Type-II 2HDM.}\label{tab}
\end{table*}

From~Table~\ref{tab}, we observe that there are two ways to realise
the SM alignment limit:
\begin{itemize}
\item SM-like $h$ scenario: 
$M_h \approx \text{125 GeV},\,  \sin(\beta-\alpha) = 1 ,\,\,  \text{with} \,\,  \beta-\alpha=\pi/2$.
 \item SM-like $H$ scenario:
$M_H \approx \text{125 GeV},\,
\cos(\beta-\alpha) = 1,\,\,  \text{with} \,\, \beta=\alpha$.
\end{itemize}
In these limits, the CP-even $H\, (h)$ scalar couples to the EW gauge
bosons with coupling strength exactly as that of the SM Higgs boson,
while $h\, (H)$ does not couple to them at all~\cite{Dev:2014yca}.
In the above two scenarios, the SM-like Higgs boson is identified with
the $125$-GeV resonance observed at the LHC \cite{Aad:2012tfa,Chatrchyan:2012xdj}.  In the
literature, the neutral Higgs partner ($H$) in the SM-like $h$
scenario is usually termed the heavy Higgs boson. Instead, in the
SM-like $H$ scenario, the partner particle $h$ can only have a mass
smaller than $\sim$\,125 GeV ~\cite{Bernon:2015qea}.  In this paper, we consider the
alignment limit with $\beta=\alpha$, which falls in the category of the
SM-like $H$ scenario, but the CP-even scalar partner $h$ can be either
lighter or heavier than the observed scalar resonance at the LHC.  In
the alignment limit, the SM-like Higgs boson becomes aligned
with one of the neutral eigenstates.

In the so-called Higgs basis \cite{Georgi:1978ri}, the CP-even mass matrix
$M_S^2$ given in~(\ref{abc}) takes on the form
 \begin{eqnarray}
\widehat{M}_S^2 & = & \left(
 \begin{matrix}
 c_{\beta} & s_{\beta} \\
 -s_{\beta} & c_{\beta}
 \end{matrix}
 \right) M_S^2 \left(
 \begin{matrix}
 c_{\beta} & -s_{\beta} \\
 s_{\beta} & c_{\beta}
 \end{matrix}
 \right)= \left(
 \begin{matrix}
 \widehat{A} & \widehat{C} \\
 \widehat{C} & \widehat{B}
 \end{matrix}
 \right),
\end{eqnarray}
with
\begin{eqnarray}
\widehat{A} &=& 2v^2 \left[ c_{\beta}^4 \lambda_1 + s_{\beta}^2 c_{\beta}^2 \lambda_{345}
 + s_{\beta}^4 \lambda_2 + 2 s_{\beta} c_{\beta} \left( c_{\beta}^2 \lambda_6 
 + s_{\beta}^2 \lambda_7 \right) \right], 
\nonumber \\
\widehat{B} &=& M_a^2 + \lambda_5 v^2 + 2v^2 \left[ s_{\beta}^2 c_{\beta}^2 \left(
 \lambda_1 + \lambda_2 - \lambda_{345} \right) - s_{\beta} c_{\beta} \left(
 c_{\beta}^2 - s_{\beta}^2 \right) \left(\lambda_6 - \lambda_7 \right) \right],
 \\ 
\widehat{C} &=& v^2 \left[ s_{\beta}^3 c_{\beta} \left( 2 \lambda_2 - \lambda_{345} \right)
 - c_{\beta}^3 s_{\beta} \left( 2 \lambda_1 - \lambda_{345} \right)
 + c_{\beta}^2 \left( 1 - 4 s_{\beta}^2 \right) \lambda_6
 + s_{\beta}^2 \left(4 c_{\beta}^2 - 1 \right) \lambda_7 \right].
\nonumber 
\end{eqnarray}

The SM alignment limit, $\cos(\beta-\alpha) \to 1$, can be realised in
two different ways: (i)~$\widehat{C} \to 0$ and
(ii)~$M_{h^\pm}\!\sim\!M_a \gg v$.  The first realisation~(i) does not
depend on the choice of the non-SM scalar masses, such as $M_{h^\pm}$
and $M_a$, whereas the second one~(ii) is only possible in the
well-known decoupling
limit~\cite{Georgi:1978ri,Gunion:2002zf,CP2005}.  In
the first case, SM alignment is obtained by setting $\widehat{C} = 0$,
which in turn implies the condition~\cite{Dev:2014yca}:
 \begin{equation}
  \label{A-C}
\lambda_7 t_{\beta}^4 - \left( 2 \lambda_2 - \lambda_{345} \right) t_{\beta}^3 
+ 3 \left( \lambda_6 - \lambda_7 \right) t_{\beta}^2 
+ \left( 2 \lambda_1 - \lambda_{345} \right) t_{\beta} - \lambda_6\ =\ 0\;.
\end{equation}
Barring fine-tuning among quartic couplings, \eqref{A-C} leads to
the following constraints:
\begin{equation}
   \label{eq:NAcond}
\lambda_1\ =\ \lambda_2\ =\ {\lambda_{345} \over 2}\;, \qquad 
\lambda_6\ =\ \lambda_7\ =\ 0\;,
\end{equation}
which are independent of $\tan\beta$ and non-standard scalar masses.
In this case, the two CP-even Higgs masses in the alignment limit are given by,
\begin{eqnarray}
M_H^2\ & =&\ 2 v^2 \left( \lambda_1 c_{\beta}^4 + \lambda_{345} s_{\beta}^2 c_{\beta}^2
 + \lambda_2 s_{\beta}^4 \right)\ \equiv\ 2 \lambda_{\text{SM}} v^2, \\ 
M_h^2\ &=&\ M_a^2 + \lambda_5 v^2 + 2v^2 s_{\beta}^2 c_{\beta}^2 \left(
 \lambda_1 + \lambda_2 - \lambda_{345} \right).
\end{eqnarray}

In the second realisation~(ii) mentioned above, we may simplify
matters by expanding $M^2_{H,h}$ in powers of $v/M_a\ll 1$. In this
way, we obtain~\cite{Dev:2014yca}
\begin{eqnarray}
M_H^2\ &\simeq&\ 2\lambda_{\text{SM}} v^2 - {v^4 s_{\beta}^2 c_{\beta}^2
               \over M_a^2 + \lambda_5 v^2} \,
 \Big[ s_{\beta}^2 \left( 2 \lambda_2 - \lambda_{345} \right)
 - c_{\beta}^2 \left( 2 \lambda_1 - \lambda_{345} \right) \Big]^2\,,\\
M_h^2\ &\simeq&\ M_a^2 + \lambda_5 v^2\ \gg\ v^2\; .
\end{eqnarray}
Note that at large $\tan\beta$, the phenomenological properties of the
$H$-boson resemble more and more those of the SM Higgs
boson~\cite{Delgado:2013zfa,Carena:2013ooa}. Since we are interested in
analysing the deviation of the $H$-boson couplings from their SM
values, we follow an approximate approach inspired by the seesaw
mechanism~\cite{Minkowski:1977sc}. In particular, we may express all
the $H$-boson couplings in terms of the light-to-heavy scalar-mixing
parameter $\widehat{C}/\widehat{B}$.  Thus, employing~\eqref{alpha}
for the hatted quantities and ignoring $\widehat{A}$ next to
$\widehat{C}$, we may derive the approximate analytic expressions
\begin{subequations} \label{ex-gc}
\begin{align}
g_{HVV}&\simeq 1-{\widehat{C}^2\over 2 \widehat{B}^2}\;,
\\
g_{hVV}&\simeq -{\widehat{C}\over \widehat{B}}\ =\ 
{v^2 s_{\beta} c_{\beta} \over M_a^2 + \lambda_5 v^2}\,
 \Big[\, c_{\beta}^2 \left( 2 \lambda_1 - \lambda_{345} \right) -
         s_{\beta}^2 \left( 2 \lambda_2 - \lambda_{345} \right)
         \Big]\;. 
\end{align}
\end{subequations}
Given the tight experimental limits on the deviation of $g_{HVV}$
from~1, one must have that the light-to-heavy scalar mixing parameter
$\widehat{C}/\widehat{B} \ll 1$, which justifies our seesaw-inspired
approximation. In fact, in the exact SM alignment limit,
$\alpha \to \beta$, the mixing parameter $\widehat{C}/ \widehat{B}$
vanishes identically.  

In similar fashion, we may derive approximate analytic expressions for the 
$h$- and $H$-boson couplings to up- and down-type quarks.
To leading order in the light-to-heavy scalar
mixing~$\widehat{C}/\widehat{B}$, these are given by
\begin{subequations} 
    \label{mfc}
\begin{align}
 g_{huu}&\simeq -{\widehat{C}\over \widehat{B}}+\frac{1}{\tan\beta}\;,\\ 
 g_{hdd}&\simeq -{\widehat{C}\over \widehat{B}}-\tan\beta\;,\\ 
 g_{Huu}&\simeq  1+\frac{1}{\tan\beta}\,{\widehat{C}\over \widehat{B}}\;,\\ 
 g_{Hdd}&\simeq 1-{\widehat{C}\over \widehat{B}}\,\tan\beta\;.
\end{align}
\end{subequations}
In the SM alignment limit, we have $g_{Huu} \to 1$ and
$g_{Hdd} \to 1$.  Obviously, any deviation of the $g_{Huu}$ and
$g_{Hdd}$ couplings from their SM values is
controlled by $\tan \beta$ and $\widehat{C}/\widehat{B}$.

In the present study, our primary interest lies in natural
realisations of SM alignment, for which neither a mass hierarchy
$M_a\gg v$, nor a fine-tuning among the quartic couplings will be
necessary.  To this end, one is therefore compelled to identify
possible maximal symmetries of the 2HDM potential that would impose
the condition stated in~\eqref{A-C}. 
In~the next section, we will show how SM alignment can be
achieved naturally by virtue of an SO(5) symmetry imposed on the theory.

\section{The Maximally Symmetric 2HDM } \label{maxsym}

A convenient field basis to describe the 2HDM potential will be to make
use of an 8-dimensional SU(2)$_L$-covariant multiplet
representation~\cite{Battye:2011jj,Nishi:2011gc},  
\begin{equation}
  \label{eq:bfPhi} 
\bf{\Phi}=\left(
 \begin{matrix}
\Phi_1 \\ \Phi_2 \\ \widetilde{\Phi}_1 \\ \widetilde{\Phi}_2\\
 \end{matrix}\right)\,,
\end{equation}
with $\widetilde{\Phi}_{1,2} = i\sigma_2 \Phi^*_{1,2}$.  With the help
of the multiplet ${\bf \Phi}$, a $6$-dimensional Lorentz vector
$R^A = \Phi^\dagger \Sigma^A \Phi$ may be constructed, where
$A = 0,\, 1,\dots,\, 5$ and $\Sigma^A$ are $8\times 8$-dimensional
matrices whose precise analytic form is given
in~\cite{Battye:2011jj,Pilaftsis:2011ed}. In this bilinear field-space
formalism, $R^A$ transforms under the {\em orthochronous} SO(1,5)
symmetry group~\cite{Dev:2014yca,Pilaftsis:2011ed}. Since the
SU(2)$_L$ gauge-kinetic terms of the scalar doublets $\Phi_{1,2}$ must
be canonical, the set of SO(1,5) rotations reduces to those of SO(5).
So, in the absence of the hypercharge gauge coupling $g'$ and fermion
Yukawa couplings, the maximal symmetry group of the 2HDM is SO(5).  As
a consequence, the 2HDM potential has a total of 13 accidental
symmetries~\cite{Battye:2011jj}, of which 6 preserve
U(1)$_Y$~\cite{Ivanov:2007de,Ferreira:2009wh,Ferreira:2010yh} and 7
are custodially symmetric~\cite{Pilaftsis:2011ed}.  Given the
isomorphism of the Lie algebras: SO(5) $\sim$ Sp(4), the maximal
symmetry group of the 2HDM in the original $\bf{\Phi}$-field space is
$G^{\bf{\Phi}}_{\text{2HDM}} = [\text{Sp(4)}/\text{Z}_2] \times
\text{SU(2)}_L$~\cite{Pilaftsis:2011ed}.
In the Type-II 2HDM, the maximal symmetry group SO(5) is the simplest
of the three possible symmetries that can realise natural SM
alignment. The other two are in bilinear [original] field
space~\cite{Dev:2014yca}: (i) O(3) $\otimes$ O(2) [SU(2)$_{\rm HF}$]
and (ii) Z$_2 \otimes$ [O(2)$]^2$ [SO(2)$_{\rm HF} \times { CP}$].  In
what follows, we focus on the simplest realisation of SM alignment,
which is called the MS-2HDM.

In the MS-2HDM, the SO(5) symmetry puts severe restrictions on the allowed
form of the kinematic parameters of the 2HDM
potential in~\eqref{eq:V2HDM},
\begin{eqnarray}
\label{m-c}
\mu_1^2 &=& \mu_2^2\,, \qquad m_{12}^2 = 0\,, \nonumber \\
\lambda_2 &=& \lambda_1\,, \qquad\ \ \lambda_3 = 2\lambda_1\,, \qquad 
\lambda_4 = \text{Re}(\lambda_5) = \lambda_6 = \lambda_7 = 0\;.
\end{eqnarray}
Evidently, the MS-2HDM potential obeys naturally the alignment constraints
given in~\eqref{eq:NAcond}. As a consequence, the scalar potential takes
on a very simple form,
\begin{equation}
   \label{eq:VMS2HDM}
V\ =\ - \mu^2 \left( \left| \Phi_1 \right|^2\: +\: \left| \Phi_2 \right|^2 \right) 
 + \lambda \left( \left| \Phi_1 \right|^2 + \left| \Phi_2 \right|^2
      \right)^2\ +\ \Delta V,
\end{equation}
where
\begin{equation}
   \label{eq:Vsoft}
\Delta V\ =\ \displaystyle\sum_{i,j=1,2}^{i\neq j}\, m_{ij}^2\, ( \Phi_i^{\dagger} \Phi_j)
\end{equation}
are soft SO(5)-breaking mass terms, which are introduced here for
phenomenological reasons as we will explain below.

After EW symmetry breaking, the following breaking pattern emerges:
\begin{equation}
\text{SO(5)}\ \xrightarrow[]{\left< \Phi_{1,2} \right> \neq 0}\ \text{SO(4)}.
\end{equation}
If $\Delta V = 0$, the CP-even scalar $H$ receives a non-zero squared mass
$M^2_H= 2 \lambda_2 v^2$, while the other scalars, $h,\, a$ and
$h^\pm$, are all massless.  These are massless pseudo-Goldstone bosons
that have sizeable couplings to the SM gauge bosons. Accordingly,
several experimentally excluded decay channels would open,
e.g.~$Z \to h a$ and $W^\pm \to h^\pm h$~\cite{22}.  If the SO(5)
symmetry is realised at some high energy scale $\mu_X$ (much above the
EW scale), then due to RG effects the following breaking pattern may
take place~\cite{Dev:2014yca}:
\begin{eqnarray}
\text{SO(5)} \otimes \text{SU(2)}_L &\xrightarrow[]{g' \neq 0}&
 \text{O(3)} \otimes \text{O(2)} \otimes \text{SU(2)}_L 
 \sim \text{O(3)} \otimes \text{U(1)}_Y \otimes \text{SU(2)}_L 
 \nonumber \\ &\xrightarrow[]{\text{Yukawa}}&
 \text{O(2)} \otimes \text{U(1)}_Y \otimes \text{SU(2)}_L
 \sim \text{U(1)}_{\text{PQ}} \otimes \text{U(1)}_Y \otimes \text{SU(2)}_L
 \nonumber \\ &\xrightarrow[]{\left< \Phi_{1,2} \right> \neq 0}&
 \text{U(1)}_{\text{em}}.
\end{eqnarray}
Note that the RG effect of the gauge coupling $g^{\prime}$ only lifts
the charged Higgs mass $M_{h^\pm}$, while the corresponding effect of
the Yukawa couplings (particularly that of the top-quark~$y_t$)
renders the other CP-even pseudo-Goldstone boson~$h$ massive. Instead,
the CP-odd scalar $a$ remains massless and can be identified with a
Peccei--Quinn~(PQ) axion after the SSB of a global U(1)$_{\rm PQ}$
symmetry~\cite{Peccei:1977hh,Weinberg:1977ma,Wilczek:1977pj}.  Since
weak-scale PQ axions have been ruled out by experiment, we have
allowed for the SO(5) symmetry of the MS-2HDM potential
in~\eqref{eq:VMS2HDM} to be broken by the soft SO(5)-breaking mass
terms in~\eqref{eq:Vsoft}. Hence, without loss of generality, we may
consider that only the soft SO(5)-breaking parameter Re$(m^2_{12})$ is
non-zero. With this minimal addition to the MS-2HDM potential, the
scalar-boson masses are given, to a very good approximation, by
\begin{equation}
    \label{eq:m12}
 M_H^2 = 2\lambda_2 v^2, \qquad M_h^2\ =\ M_a^2\ =\ M_{h^{\pm}}^2\ 
 =\ {\text{Re}(m_{12}^2) \over s_{\beta} c_{\beta}}\; .
\end{equation}
Hence, all pseudo-Goldstone bosons, $h$, $a$ and $h^\pm$,  become
massive and almost degenerate in mass. As we will see in the next
section, this degeneracy is very stable against RG effects even up to
two-loop order.

In our study, we will consider the charged Higgs boson mass
$M_{h^{\pm}}$ as an input parameter above 500-GeV range, in agreement with $B$-meson constraints \cite{Misiak:2017bgg}. It will also be our threshold
above which all parameters run with 2HDM renormalisation group
equations (RGEs). Moreover, we implement the matching conditions with
two-loop RG effects of the SM at given $M_{h^{\pm}}$ threshold scales.
We also employ two-loop 2HDM RGEs to find the running of the gauge,
Yukawa and quartic couplings at RG scales larger
than~$M_{h^{\pm}}$. For reviews on one-, two- and three-loop RGEs in
the 2HDM,
see~\cite{Dev:2014yca,Oredsson:2018yho,Chowdhury:2015yja,Krauss:2018thf,Bednyakov:2018cmx}.
The SM and the 2HDM RGEs have been computed using the public
\texttt{Mathematica} package \texttt{SARAH} \cite{Sarah}, which has
been appropriately adapted for the MS-2HDM.

\section{Quartic Coupling Unification} \label{QCU}

As we saw in the previous section, the SO(5) symmetry of the
MS-2HDM potential is broken explicitly by RG effects and soft-mass terms.  In
this section, we will consider a unified theoretical\- framework in
which the SO(5) symmetry is realised at some high-energy
scale~$\mu_X$, where all the conditions in~\eqref{m-c} are met.
Of particular interest is the potential existence of conformally-invariant 
unification points at which all quartic couplings of the MS-2HDM potential
vanish simultaneously.

To address the above issue of quartic coupling unification, we employ
two-loop RGEs for the MS-2HDM from the unification scale $\mu_X$ to
the charged Higgs-boson mass~$M_{h^{\pm}}\ll \mu_X$. Below this
threshold scale $\mu_{\rm thr} = M_{h^{\pm}}$, the SM is a good
effective field theory, so we use the two-loop SM RGEs given
in~\cite{Carena:2015uoe} to match the relevant MS-2HDM couplings to
the corresponding SM quartic coupling $\lambda_{\rm SM}$, the Yukawa
couplings, the SU(2)$_L$ and U(1)$_Y$ gauge couplings $g_2$ and $g'$
(with $g_1 = \sqrt{5/3}g'$).  In order to obtain illustrative
predictions, we have chosen our threshold scales to be at
$\mu_{\rm thr}\ =\ M_{h^{\pm}}\ =\ 500$~GeV, 1~TeV, 10~TeV and
100~TeV.

The theoretical SM parameters are determined in terms of precision
observables, such as the $Z$-boson mass ($M_Z$), the $W^\pm$-boson
mass ($M_W$), the Fermi constant $G_\mu$, the strong coupling
$\alpha_3(M_Z)$, the top-quark mass ($M_t$) and the Higgs boson mass
($M_{H_{\rm SM}}$). In detail, the following SM values will be adopted~\cite{Buttazzo:2013uya}:
\begin{align}
\quad &M_W= 80.384 \pm 0.014 \text{ GeV,} &&M_Z= 91.1876 \pm 0.0021
                                             \text{ GeV,} \nonumber \\[0.05in] 
\quad &M_{H_{\rm SM}}= 125.15 \pm 0.24 \text{ GeV,} &&M_t= 173.34 \pm 0.76 \pm
                                            0.3 \text{ GeV,}  \nonumber \\[0.05in]   
\quad &v= (\sqrt{2}G)^{-1/2}= 246.21971 \pm 0.00006 \text{ GeV,}
                                          &&\alpha_3(M_Z) = 0.1184 \pm 0.0007.  
\end{align}
The values of the two-loop SM couplings at different threshold
scales, $\mu_{\rm thr} = M_t,\, M_{h^{\pm}}$, are given in Table~\ref{tab1}.

\begin{table*}[t]
 \centering
 \small
\begin{tabular}{ c c c c c c c c c c c c c c c c c c c c c c c c c c }
 \hline \hline
$\mu_{\rm thr}$ [GeV]&&& $g_1$ &&& $g_2$ &&& $g_3$ &&& $\lambda_{\rm SM}$ &&& $y_t$  &&& $y_b$ &&&$y_{\tau}$ &&\\ \hline
\, $M_t$  &&& 0.3583 &&& 0.6779 &&& 1.1666 &&& 0.1292  &&& 0.9401 &&& 0.0157 &&& 0.01000 && \\ 
\, $500$  &&& 0.3605 &&& 0.6423 &&& 1.0910 &&& 0.1102  &&& 0.8807 &&& 0.0146 &&& 0.01008 && \\ 
\, $10^3$ &&& 0.3619 &&& 0.6388 &&& 1.0523 &&& 0.0997  &&& 0.8521 &&& 0.0140 &&& 0.01012 &&\\
\, $10^4$ &&& 0.3668 &&& 0.6274 &&& 0.9484 &&& 0.0719  &&& 0.7746 &&& 0.0123 &&& 0.01022 &&\\
\, $10^5$ &&& 0.3718 &&& 0.6164 &&& 0.8704 &&& 0.0518  &&& 0.7154 &&& 0.0111 &&& 0.01026 &&\\ \hline \hline 
\end{tabular}
\caption{\it SM couplings evaluated at two-loop level in the
  $\overline{\rm MS}$ scheme at various threshold scales~$\mu_{\rm thr}$, 
  i.e.~for $\mu_{\rm thr} = M_t$
  and $\mu_{\rm thr} = M_{h^{\pm}} = 500$~GeV, 1~TeV, 10~TeV and 100~TeV.}\label{tab1} 
\end{table*}

The matching conditions for the Yukawa couplings at the threshold
scale~$\mu_{\rm thr} = M_{h^\pm}$ read
\begin{equation}
   \label{eq:Ymatch}
 h_t^{\text{MS-2HDM}} = {y_t \over s_{\beta}}\;, \qquad
h_b^{\text{MS-2HDM}} = {y_b \over c_{\beta}}\;, \qquad
h_{\tau}^{\text{MS-2HDM}} = {y_{\tau} \over c_{\beta}}\;.
\end{equation}
Note that for RG scales $\mu > \mu_{\rm thr} = M_{h^{\pm}}$, the
evolution of the Yukawa couplings $h_t$, $h_b$ and $h_\tau$ is
governed by two-loop 2HDM RGEs.

\begin{figure}[!t]
\tiny
\includegraphics[width=0.8\textwidth]{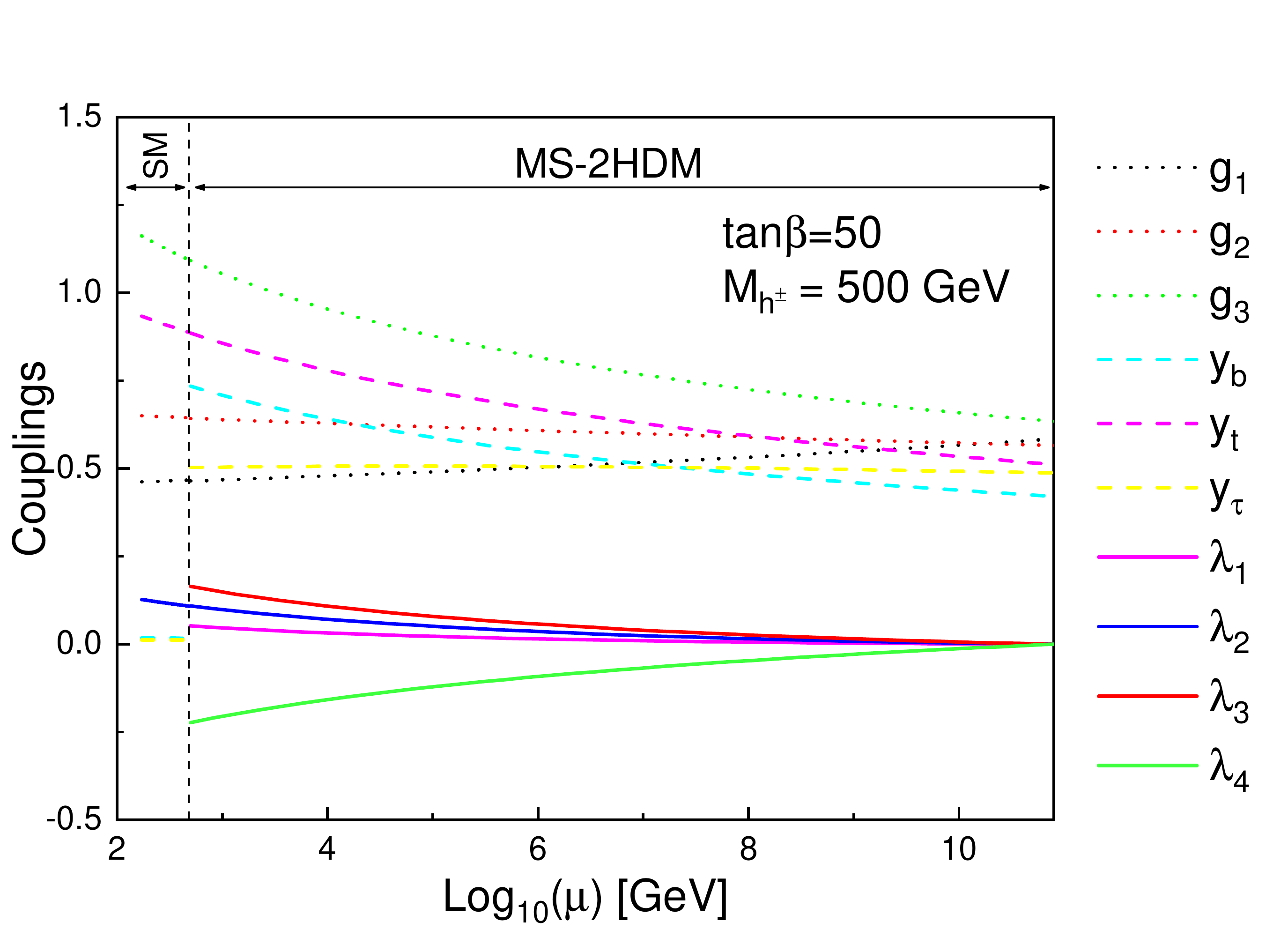}
\caption{\it The RG evolution of the quartic couplings
  $\lambda_{1,2,3,4}$, gauge and Yukawa couplings from the threshold
  scale $M_{h^\pm} = 500$ GeV up to the  their first quartic coupling unification
  scale~${\mu^{(1)}_X=10^{11}}$\,GeV for $\tan\beta = 50$.}
\label{0.5TeV}
\end{figure}
\begin{figure}[!h]   
\tiny
\includegraphics[width=0.8\textwidth]{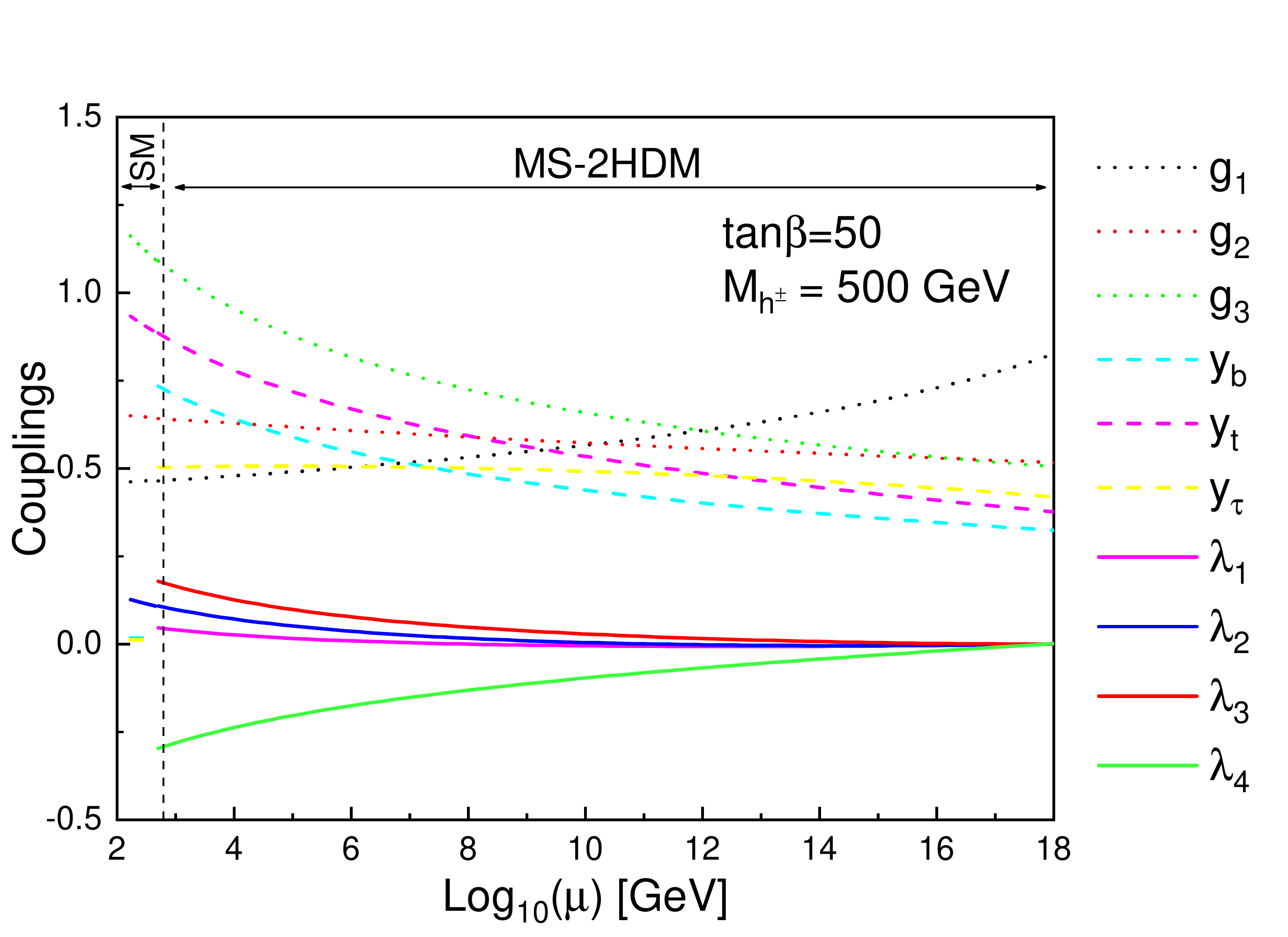}
\caption{\it The same as in Fig.~\ref{0.5TeV}, where the RG evolution
  is extended up to the second quartic coupling unification point
  $\mu^{(2)}_X=10^{18}$\,GeV is shown.}
\label{0.5TeV2}
\end{figure}

Figures \ref{0.5TeV} and \ref{0.5TeV2} display the RG evolution of all
relevant couplings of the SM and the MS-2HDM, for $\tan \beta=50$ and
$m^2_{12}\approx 70^2$ GeV$^2$ ($M_{h^\pm} = 500$~GeV). The vertical dashed line indicates the
threshold scale $\mu_{\rm thr} = M_{h^\pm} = 500$~GeV. At this RG
scale,  we observe significant discontinuities in the running of Yukawa couplings $y_b$ and
$y_\tau$ due to the matching conditions in~\eqref{eq:Ymatch}.

As can be seen from Figure~\ref{0.5TeV}, the quartic coupling
$\lambda_2$, which determines the SM-like Higgs-boson mass~$M_H$,
decreases at high RG scales due to the running of the top-Yukawa
coupling $h_t$ and turns negative just above the quartic coupling
unification scale $\mu_X \sim 10^{11}$\,GeV, at which all quartic
couplings vanish.  Thus, for energy scales above the RG scale $\mu_X$,
we envisage that the MS-2HDM will need to be embedded into another
UV-complete theory. Nevertheless, according to our estimates
in~Section~\ref{vac}, we have checked that the resulting MS-2HDM
potential leads to a metastable but sufficiently long-lived EW vacuum,
whose lifetime is many orders of magnitude larger than the age of our
Universe. In this respect, we regard the usual constraints
derived from convexity conditions on 2HDM potentials ~\cite{Deshpande:1977rw}
to be over-restrictive and unnecessary in our theoretical framework.

Of equal importance is a second conformally-invariant unification
point $\mu^{(2)}_X$ at energy scales close to the reduced Planck mass of
order $10^{18}$\,GeV, as shown in Figure~\ref{0.5TeV2}. In this case,
the key quartic coupling~$\lambda_2$ increases and becomes positive
again. Hence, in this second class of settings, any embedding of the MS-2HDM into
a UV-complete theory must have to take quantum gravity into account as well.

\begin{figure}[t]
 \tiny
\includegraphics[width=0.7\textwidth]{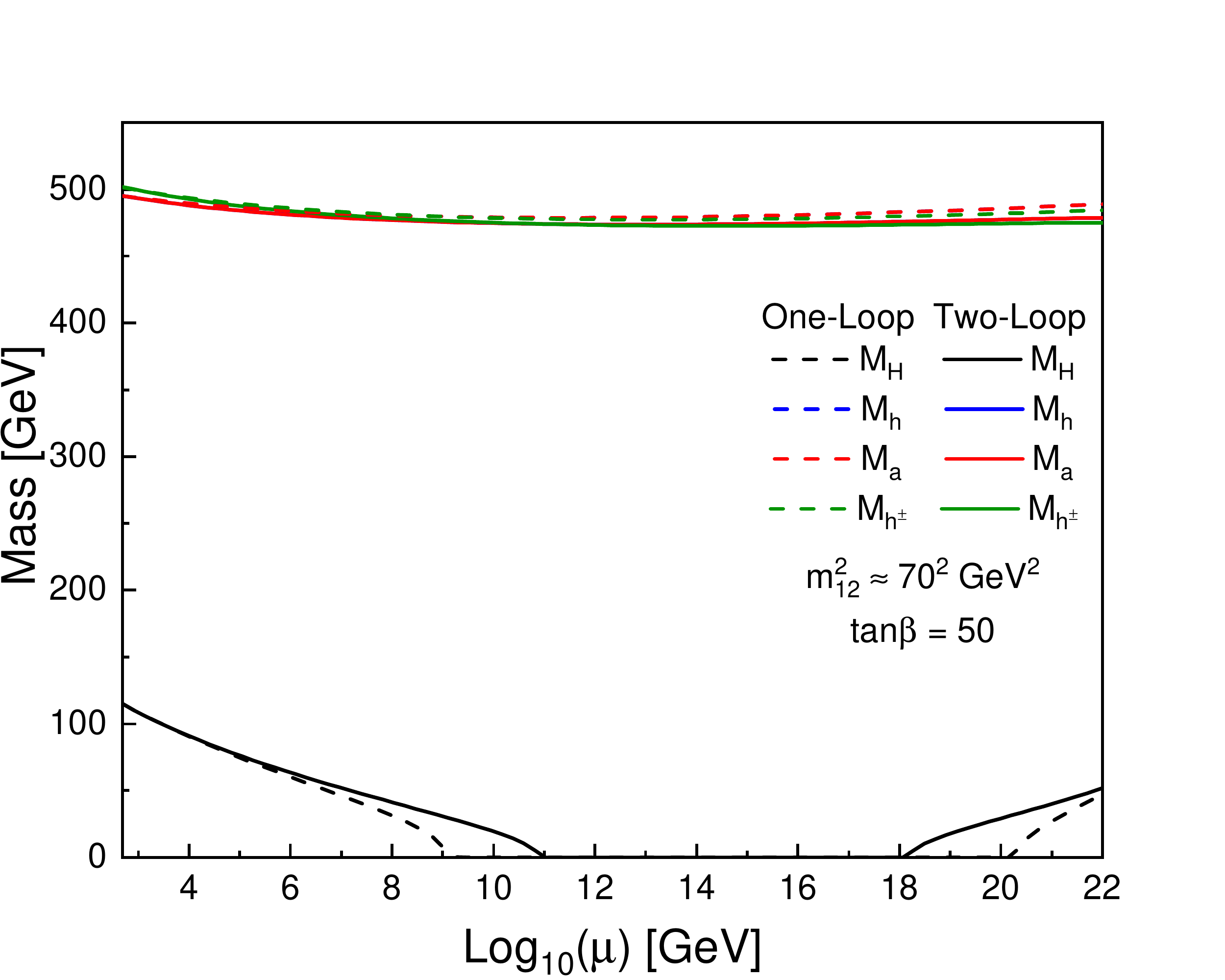}
\caption{\it The scalar mass spectrum of the MS-2HDM with $m^2_{12}
  \approx 70^2$\,GeV$^2$ and $\tan\beta = 50$. The dashed and the solid
  curves show the one-loop and the two-loop RGEs of the scalar masses,
  respectively.} 
\label{mass}
\end{figure}

In Figure \ref{mass}, we give numerical estimates of the mass spectrum
of the MS-2HDM, for $\tan \beta=50$ and $m^2_{12}\approx 70^2$ GeV$^2$
($M_{h^{\pm}}=500$ GeV).  We find that all heavy Higgs bosons, $h$,
$a$ and $h^\pm$, are approximately degenerate in mass, almost
independently of the quartic coupling unification scale~$\mu_X$. In
fact, to leading order, all masses are determined by the soft mass
term $m^2_{12}$ in~\eqref{eq:m12}.  The curve corresponding to the
SM-like Higgs-boson mass~$M_H$ is mainly tracking $\lambda_2(\mu )$, which
has two roots, one at RG scales around $\mu^{(1)}_X \sim 10^{11}$\,GeV and
another one at $\mu^{(2)}_X \sim 10^{18}$\,GeV. The latter is the largest
possible UV scale of the MS-2HDM.

\begin{figure}[t]
\tiny
\includegraphics[width=0.7\textwidth]{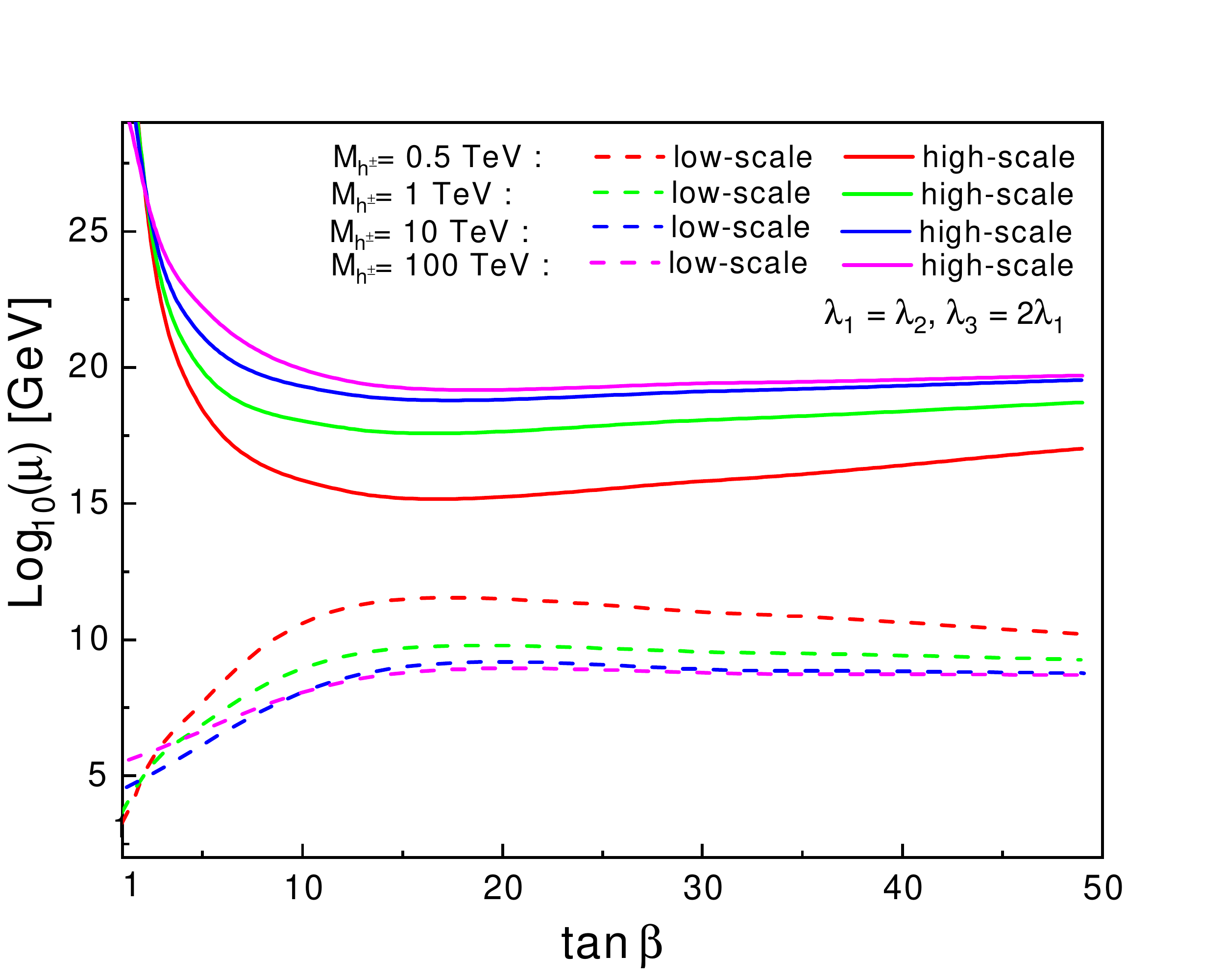}
\caption{\it Sets of quartic coupling unification points in the
  $(\tan\beta,\, \log_{10}\mu )$ plane, for charged Higgs-boson masses
  $M_{h^{\pm}}=500$~GeV, 1~TeV, 10~TeV and 100~TeV.  The dashed and
  solid curves show the sets of low-scale and high-scale quartic
  coupling unification points, $\mu^{(1)}_X$ and
$\mu^{(2)}_X$, respectively.}
\label{mix}
\end{figure}

Figure~\ref{mix} shows all conformally-invariant quartic coupling
unification points in the $(\tan\beta, \log_{10}\mu)$ plane, by
considering different values of threshold scales~$\mu_{\rm thr}$,
i.e.~for $\mu_{\rm thr} = M_{h^{\pm}}=500$\,GeV,\,$1$\,TeV,\,$10$\,TeV
and\,$100$\,TeV. The lower curves (dashed curves) correspond to sets
of low-scale quartic coupling unification points, while the upper
curves (solid curves) give the corresponding sets of high-scale
unification points. From~Figure~\ref{mix}, we may also observe the
domains in which the $\lambda _2$ coupling becomes negative. These
are given by the vertical $\mu$-intervals bounded by the lower and the
upper curves, for a given choice of $M_{h^\pm}$ and
$\tan\beta$. Evidently, as the threshold scale~$\mu_{\rm thr} = M_{h^\pm}$
increases, the size of the negative $\lambda _2$ domain increases and
becomes more pronounced for smaller values of $\tan \beta$.

Before concluding this section, it is important to stress that the
MS-2HDM requires only three additional input parameters: (i)~the soft
SO(5)-breaking mass parameter $m^2_{12}$ (or $M_{h^{\pm}})$, (ii)~the
ratio of VEVs $\tan\beta$, and (iii) the conformally-invariant quartic
coupling unification scale~$\mu_X$ which can only assume two discrete
values: $\mu^{(1)}_X$ and $\mu^{(2)}_X$. Once the values of these
three theoretical parameters are given, the entire Higgs sector of the
model can be determined. In the next section, we will give typical
predictions in terms of these three input parameters.

\section{Misalignment Predictions for Higgs Boson Couplings}\label{misa} 

In this section, we present numerical estimates of the predicted
deviations of the SM-like Higgs-boson couplings $HVV$ (with
$V= W^\pm,Z$), $Ht\bar{t}$ and $Hb\bar{b}$, from their respective SM
values. As was discussed in Section \ref{alignment}, these deviations
are controlled by the light-to-heavy scalar mixing parameter
$\widehat{C}/ \widehat{B}$. At the quartic coupling unification scale
$\mu_X$, the SO(5) symmetry of the MS-2HDM is fully restored and this
mixing parameter vanishes. However, as we saw in Section~\ref{maxsym},
RG effects induced by the U(1)$_Y$ gauge coupling and the Yukawa
couplings of the third generation of fermions break sizeably the SO(5)
symmetry, giving rise to a calculable non-zero value for
$\widehat{C}/ \widehat{B}$ and so to misalignment predictions for {\em
  all} $H$-boson couplings to SM particles.

\begin{figure}[t]
\tiny
\includegraphics[width=0.62\textwidth]{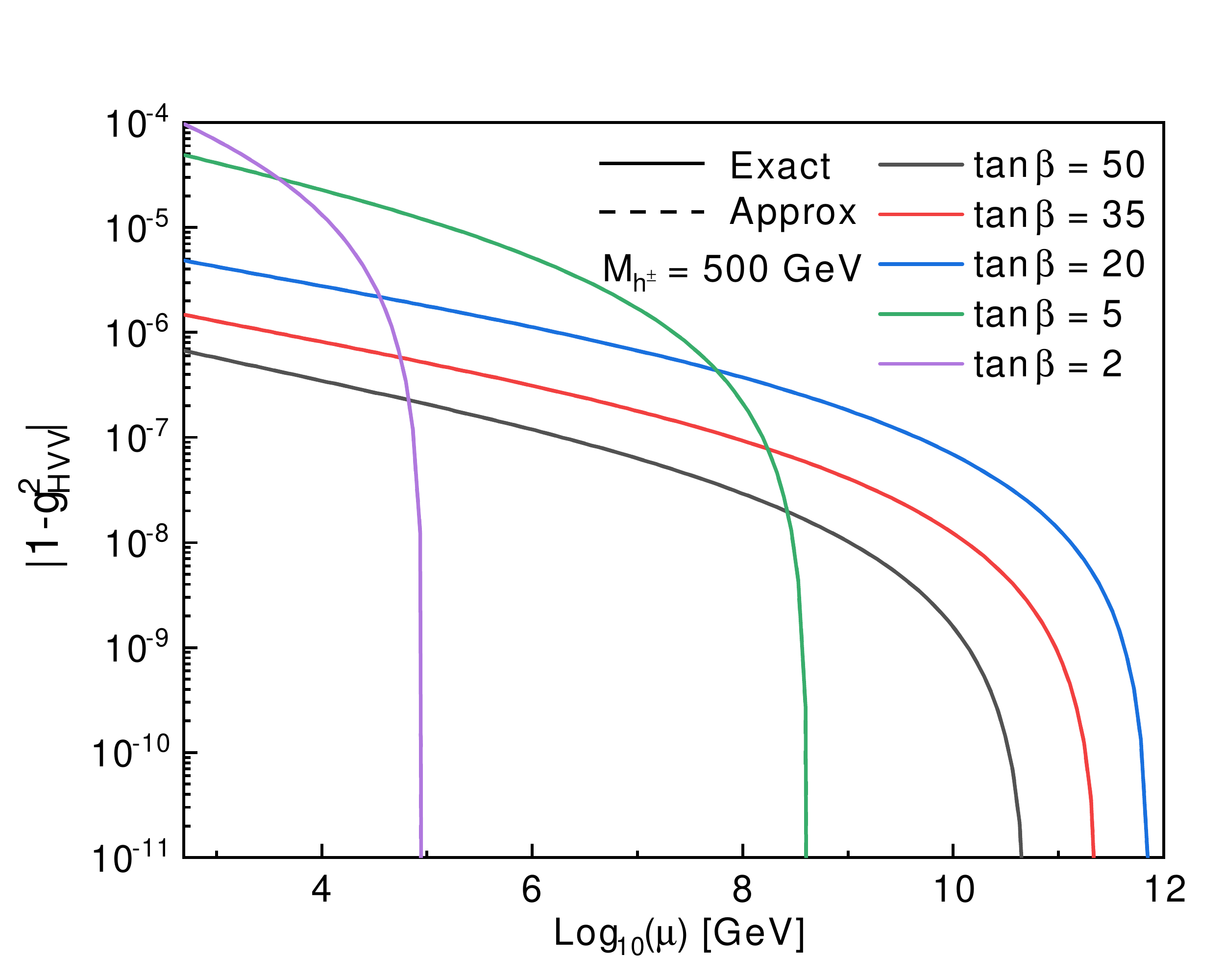}
\caption{\it Numerical estimates of the misalignment parameter $|1 -
  g^2_{HVV}|$ pertinent to the $HVV$-coupling (with $V= W^\pm,Z$) as
  functions of the RG scale~$\mu$, for a  low-scale quartic coupling
  unification scenario, assuming  $M_{h^\pm} = 500$\,GeV and
  $\tan\beta=2,\ 5,\,20,\,35$ and $50$.}
\label{1stc}
\end{figure}

\begin{figure}[h]
\tiny
\includegraphics[width=0.62\textwidth]{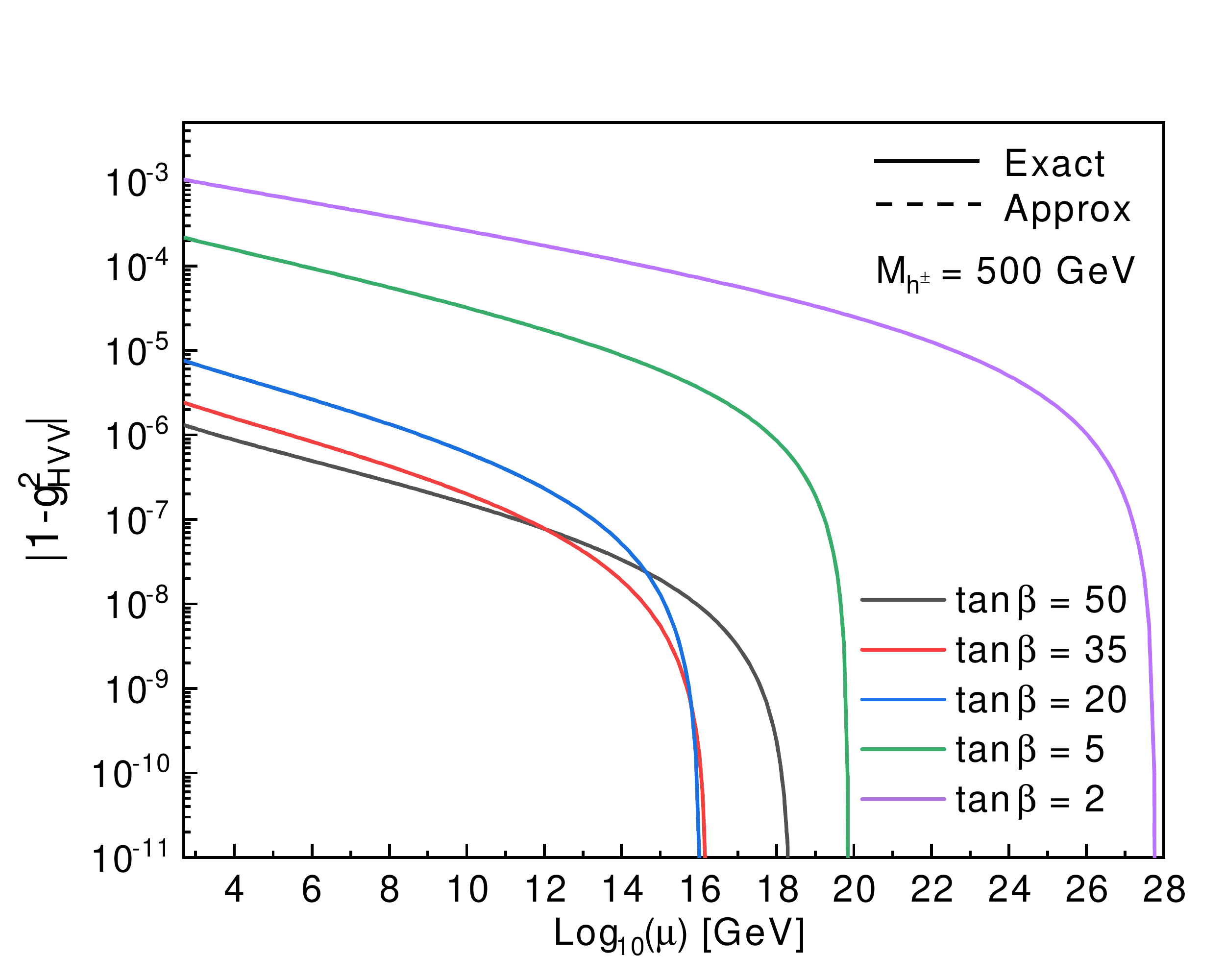}
\caption{\it The same as in Figure~\ref{1stc}, but for a high-scale quartic coupling
  unification scenario.}
\label{2ndc}
\end{figure}

In Figures~\ref{1stc} and~\ref{2ndc}, we exhibit the dependence of the
physical misalignment parameter $|1 - g^2_{HVV}|$ (with
$g_{H_{\text{SM}}V V} =1$) as functions of the RG scale~$\mu$, for
typical values of $\tan\beta$, such as $\tan\beta=2,\ 5,\,20,\,35$ and
$50$.  As expected, we observe that the normalised coupling $g_{HVV}$
approaches the SM value $g_{H_{\text{SM}}V V} =1$ at the lower- and
higher-scale quartic coupling unification points, $\mu^{(1)}_X$ and
$\mu^{(2)}_X$, as shown in Figures~\ref{1stc} and~\ref{2ndc},
respectively.  We use dashed lines to display our predictions to
leading order in $\widehat{C}/ \widehat{B}$ expansion, while solid
lines stand for the exact all-orders result. Since there is a small
deviation (below the per-mile level) of $g_{HVV}$ from the SM value,
the approximate and exact predicted values are almost overlapping.
Note that the misalignment reaches its maximum value for low values of
$\tan \beta$ and for the higher quartic coupling unification points.

\begin{figure}[h]
\tiny
\includegraphics[width=0.49\textwidth]{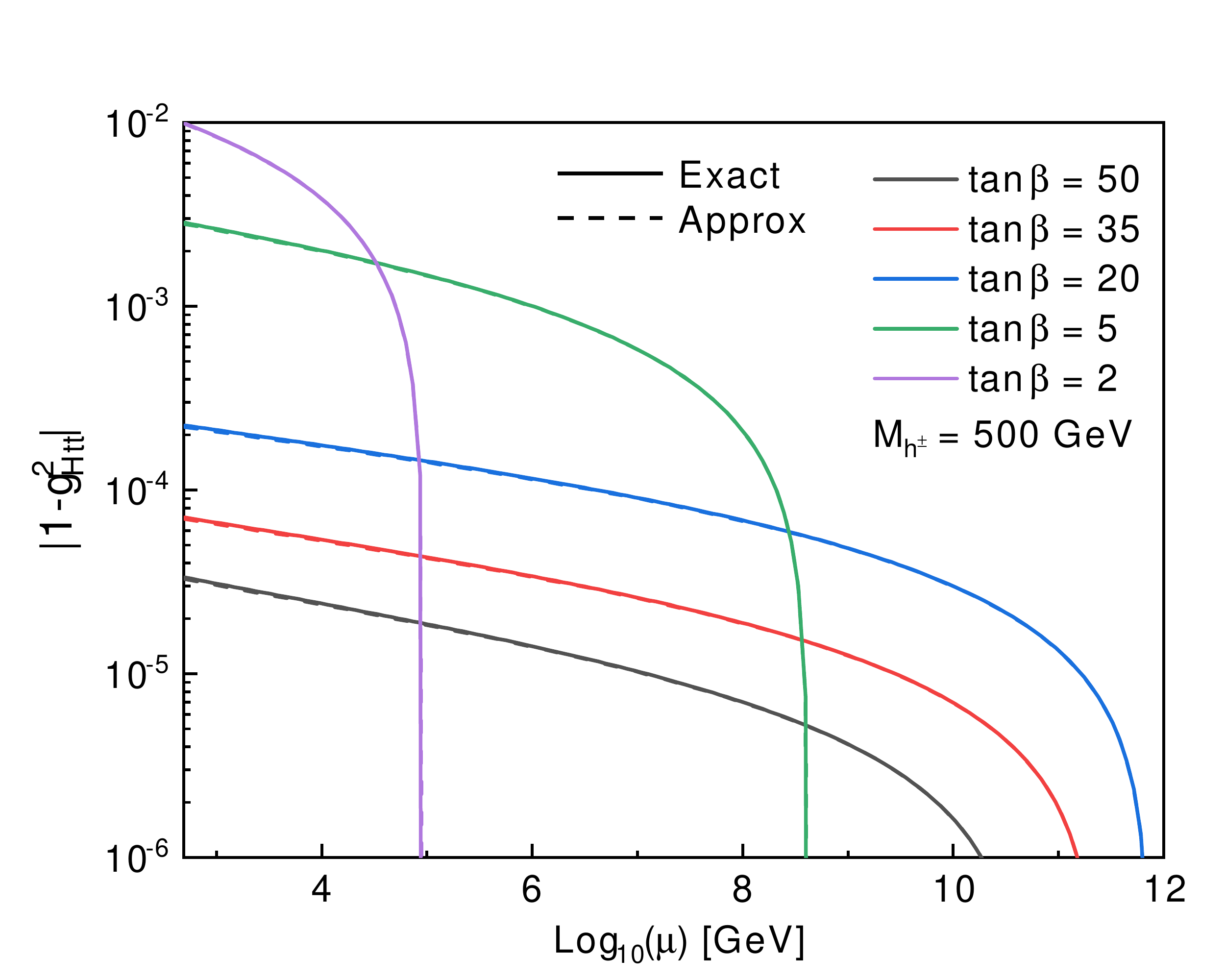}
\includegraphics[width=0.49\textwidth]{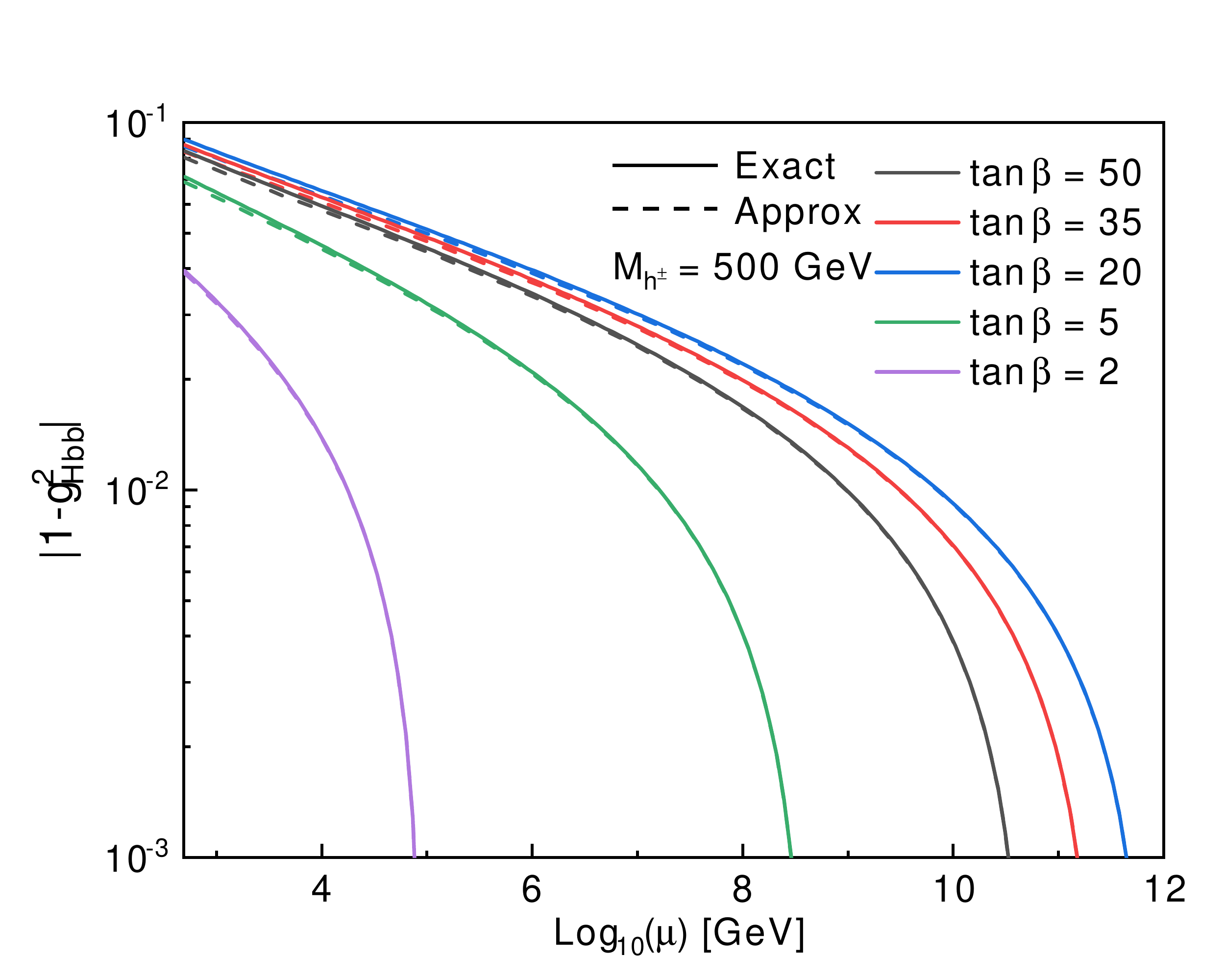}
\caption{\it The misalignment parameters $|1 -
  g^2_{Htt}|$ (left panel) and $|1 -
  g^2_{Hbb}|$ (right panel)  versus the RG scale~$\mu$, for a  low-scale quartic coupling
  unification scenario, assuming  $M_{h^\pm} = 500$\,GeV and
  $\tan\beta=2,\ 5,\,20,\,35$ and $50$.}
\label{1stcf}
\end{figure}

\begin{figure}[t]
\tiny
\includegraphics[width=0.49\textwidth]{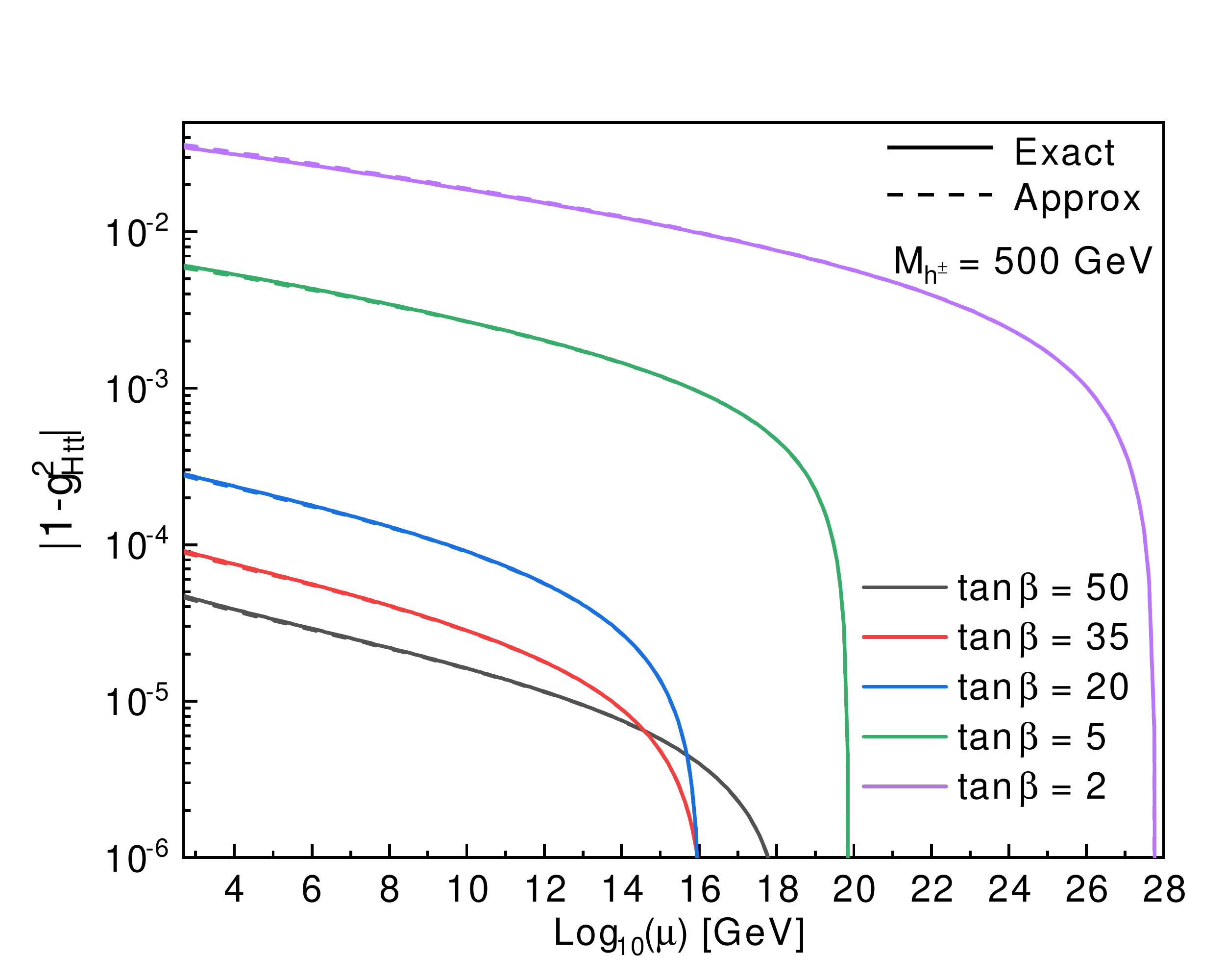}
\includegraphics[width=0.49\textwidth]{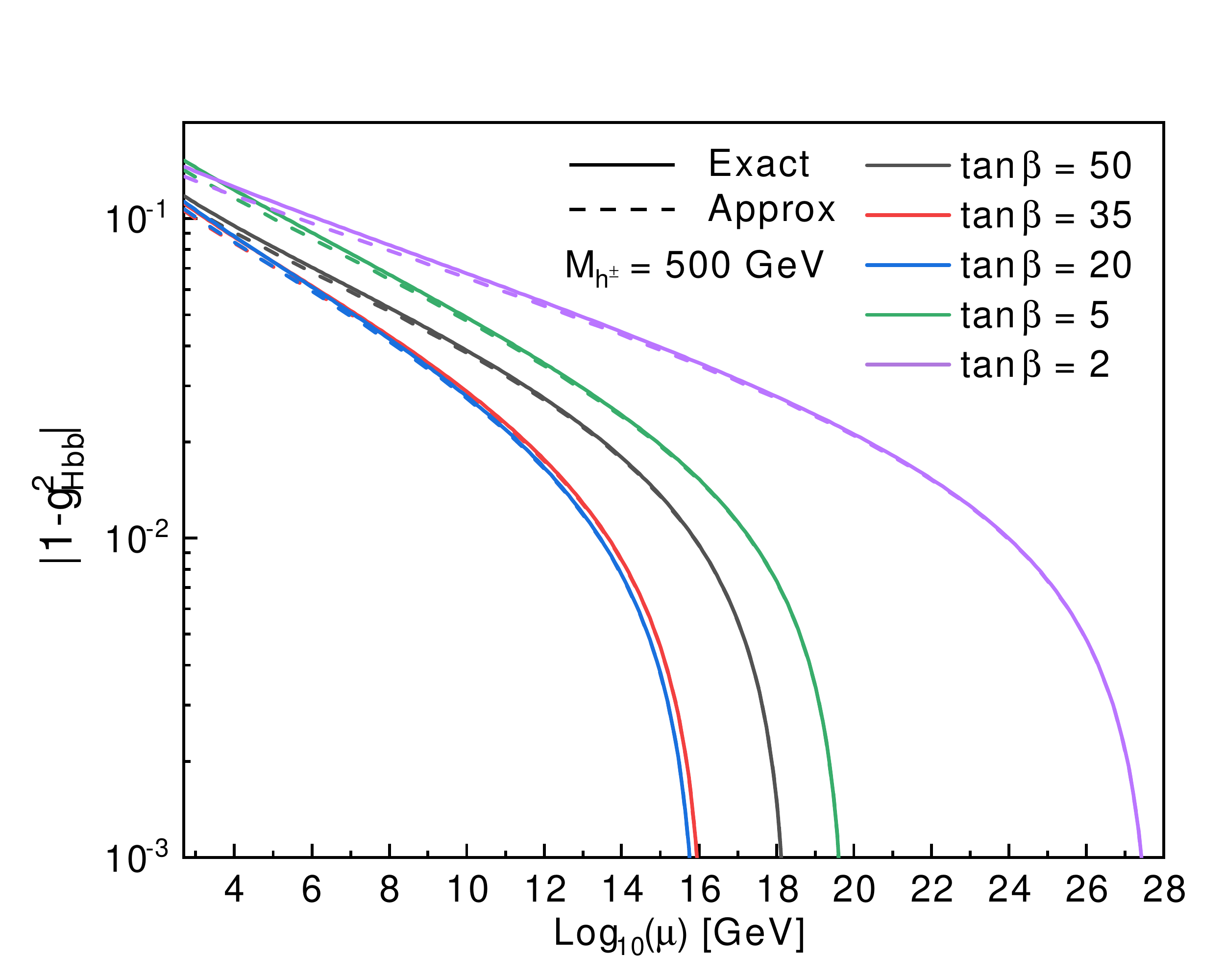}
\caption{\it The same as in Figure~\ref{1stcf}, but for a high-scale quartic coupling
  unification scenario.} 
\label{2ndcf}
\end{figure}

By analogy, Figures \ref{1stcf} and \ref{2ndcf} display misalignment
predictions for the $H$-boson couplings to top- and bottom-quarks, for
$\tan\beta=2,\ 5,\,20,\,35$ and $50$ and for lower- and higher-scale
quartic coupling unification points, respectively.  As before, the
deviation of the normalised couplings $g_{Htt}$ and $g_{Hbb}$ from
their SM values are larger for low values of $\tan \beta$,
e.g.~$\tan\beta = 2$, and for higher quartic coupling unification
points~$\mu^{(2)}_X$. This effect is more pronounced for~$g_{Hbb}$, as the degree of
misalignment might be even larger than 10\%. In this case, a comparison
between solid and dashed lines demonstrates the goodness of our
seesaw-inspired approximation in terms of~$\widehat{C}/ \widehat{B}$.

Finally, we confront our misalignment predictions for the SM-like
Higgs boson couplings, $g_{HZZ}$, $g_{Htt}$ and $g_{ Hbb}$ with
existing experimental data from ATLAS and CMS, including their
statistical and systematic uncertainties~\cite{201606}. All these
results are presented in Table \ref{ex}, for $M_{h^\pm} = 500$\,GeV.
The observed results for $g_{HZZ}$ and $g_{Htt}$ are in excellent
agreement with the SM and the MS-2HDM. Instead, the LHC data for
$g_{Hbb}$ can be fitted to the SM at the $3\sigma$ level. The latter
reduces only to $2\sigma$ in the MS-2HDM, for $\tan\beta = 2$
assuming a high-scale quartic coupling unification scenario.  Future 
precision collider experiments will be able to probe such a scenario.


\begin{table*}[h]
 \centering
 \small
 \begin{tabular}{c c c c c c c c c c c c c c c c c c c c c c c c c c }
 \hline\hline
 Couplings && ATLAS && CMS && $\tan\beta = 2$ && $\tan\beta = 5$ && $\tan\beta = 20$ && $\tan\beta = 35$ && $\tan\beta = 50$ \\ \hline
 $|g_{HZZ}^{\text{low-scale}}|$ && [0.86, 1.00]&& [0.90, 1.00] && 0.9999 && 0.9999 && 0.9999 && 0.9999 && 0.9999 \\
 $|g_{HZZ}^{\text{high-scale}}|$&& && && 0.9981 && 0.9998 && 0.9999 && 0.9999 && 0.9999 \\ \hline
 $|g_{Htt}^{\text{low-scale}}|$&& $1.31^{+0.35}_{-0.33}$ && $1.45^{+0.42}_{-0.32}$ && 1.0049 && 1.0014 && 1.0001 && 1.0000 && 1.0000 \\
 $|g_{Htt}^{\text{high-scale}}|$ && && && 1.0987 && 1.0179 && 1.0003 && 1.0001 && 1.0001 \\ \hline
 $|g_{Hbb}^{\text{low-scale}}|$ && $0.49^{+0.26}_{-0.19}$ && $0.57^{+0.16}_{-0.16}$ && 0.9803&& 0.9649 && 0.9560 && 0.9574 && 0.9590 \\
 $|g_{Hbb}^{\text{high-scale}}|$&& && && 0.8810 && 0.9264 && 0.9449 && 0.9456 && 0.9427 \\
 \hline\hline
 \end{tabular}
 \caption{\it Predicted values of the SM-like Higgs boson  couplings
   to the $Z$ boson and to top- and bottom-quarks in the MS-2HDM for both
   scenarios with low- and high-scale quartic coupling unification,
   assuming $M_{h^{\pm}}=500$\,GeV. The corresponding central values
   for these couplings from ATLAS and CMS are also given, including their
   uncertainties \cite{201606}. } 
 \label{ex}
\end{table*}

\newpage

\section{Lifetime of the Metastable Electroweak Vacuum} \label{vac}

According to our discussion in Section \ref{QCU}, we have seen that
the MS-2HDM has two conformally-invariant quartic coupling unification
points~$\mu^{(1,2)}_X$, for a given choice of the charged Higgs-boson mass
$M_{h^\pm}$ and $\tan\beta$. Typically, the first point is at
$\mu^{(1)}_X \sim 10^{11}$\,GeV and the second one at
$\mu^{(2)}_X \sim 10^{18}$\,GeV. Between these two RG scales, i.e.~for
$\mu^{(1)}_X < \mu < \mu^{(2)}_X$, the running quartic coupling
$\lambda_2(\mu )$ turns negative, which gives rise to a deeper
minimum in the effective MS-2HDM potential. This new minimum will then
be the {\em true vacuum} of the MS-2HDM.  In this case, the EW vacuum
that we currently live in becomes metastable and is usually called the
{\em false vacuum}. For the high-scale quartic coupling unification
scenario, this vacuum instability might be a problem, unless the
lifetime of the EW vacuum is much larger than the age of the Universe.

In a general 2HDM, there are three different types of ground
states~\cite{Lee1,Branco,Sher}: (i)~an EW-breaking vacuum that
preserves CP and charge (normal minimum); (ii) an EW-breaking vacuum
that breaks CP spontaneously, but keeps charge conserved; and (iii) a
charge-violating ground state, where one of the upper components
$\phi^+_{1,2}$ of the scalar doublets $\Phi_{1,2}$ acquires a non-zero
VEV. The existence of various minima in the potential may
result in tunneling between different vacua.  However, it is known
since some time~\cite{Ferreira:2004yd,Ivanov:2006yq,Branchina:2018qlf}
that if the effective 2HDM potential has at least one CP- and
charge-preserving local minimum, then any other possible
charged-violating minimum cannot be deeper than this and so tunneling
into such a local minimum will not be energetically favoured.  Hence,
our false neutral EW vacuum can only tunnel to another neutral vacuum,
and our findings are consistent with this observation in the MS-2HDM.

The probability rate $P$ for quantum tunneling through a barrier is
exponentially suppressed and it may be estimated by
\begin{equation}
P \  \sim\ e^{-\Delta S_E} \;,
\end{equation}
where $\Delta S_E$ is the Euclidean action evaluated at the $\mathbb{O}(4)$-symmetric bounce
solution~\cite{Arnold:1991cv}. The
action $\Delta S_E$ interpolates between the new phase at high field
values and the EW phase. As we will see below, $\Delta S_E$ is the
Euclidean action of the corresponding bounce minus the action of the
false vacuum configuration~\cite{11}. To~determine\- the lifetime of the
false vacuum, we have to look for the so-called bounce solutions that
satisfy the equations of motion,
\begin{equation}
{d^2 \phi_i \over dr^2}+ {3\over r}{d\phi_i \over dr}= {\partial
  V(\phi)\over \partial \phi_i}\; , 
\end{equation}
where the index $i = 1,2,\dots 5$ labels all scalar
fields~$\phi_i = (h,H,a,h^+,h^-)$ and $\phi \equiv \{\phi_i\}$. Note that
the fields $\phi_i$ depend only on the Euclidean radial coordinate
$r=(t_E^2+{\bf{x}}^2)^{1/2}$, as a consequence of the
$\mathbb{O}(4)$-symmetry of the problem.  The boundary conditions of
the equations of motion are
\begin{equation}
    \label{eq:BCbounce}   
{d\phi_i \over dr}\bigg|_{r=0}\ =\ 0\;, \,\,\,\,\,\,\,\,\,
\lim_{r\to\infty}(\phi_i)\ =\ \phi_i^{\rm fv}\;,
\end{equation} 
where $\phi_i^{\rm fv}$ are the values of the fields $\phi_i$ at the
false vacuum.  More explicitly, the action $\Delta S_{E,i}$ along a
$\phi_i$-direction is given by~\cite{Arnold:1991cv}
\begin{equation}
\Delta S_{E,i}\ \equiv\ S[\phi_i^{\rm b}]\: -\: S[\phi_i^{\rm fv}]\ =\ 
-{\pi^2\over 2} \int^{\infty}_0 dr\, r^3\, {\partial V(\phi^{\rm b}) \over
  \partial \phi^{\rm b}_i}\,\phi^{\rm b}_i\; .
\end{equation}
Here, $\phi^{\rm b} \equiv \{ \phi^{\rm b}_i\}$ denotes collectively
the bounce solutions satisfying the boundary conditions${}$ stated
in~\eqref{eq:BCbounce}.  For our MS-2HDM scenario, we have five
second-order differential equations to determine the bounce
configurations, where $r = 0$ is the center of the bounce which
asymptotically approaches the false vacuum
$\phi^{\rm fv} \equiv \{ \phi^{\rm fv}_i\}$ as $r \to \infty$.
 
As the second deeper minimum of the potential occurs at field values
$\phi _i \gg v$, where typically
$\phi_i \stackrel{>}{{}_\sim} 10^{10}$\,GeV, it will be a good
approximation if we only keep its quartic terms. Therefore, the EW
vacuum lifetime is computed by considering only the scale-invariant
part of the effective potential:
\begin{equation}
V (\mu \gg v)\ \approx\  {1\over 4}\, \Big[\,\lambda_{1}h^4+\lambda_{2}H^4+
2(\lambda_{3}+\lambda_{4})\,h^2H^2\,\Big]\;,
\end{equation}
where the running quartic coupling $\lambda_2$ is negative.  This
analytic approximation leads to the following Euclidean action for the
 Fubini instanton \cite{Fubini}:
\begin{equation} 
\Delta S_{E,i}\ \simeq\ -\, \frac{8\pi ^2}{3|\lambda_i|}\; ,
\end{equation} 
where $i = 1, 2$.  Knowing $\Delta S_{E,i}$, an approximate formula
for the EW vacuum lifetime $\tau$ in units of the age of the Universe
($T_U \sim 13.7 \times 10^9 $ years) for the individual
quantum-tunnelings from the EW vacuum to a deeper minimum in the
$\phi_i$-direction ($\phi_1= h$ and $\phi_2 = H$) may be derived,
\begin{equation}
\tau_i\ =\ {e^{\Delta S_{E,i}} \over \phi_i(0)^4\, T_U^4}\;.
\label{tau}
\end{equation}

\begin{figure}[H]
\includegraphics[width=1.0\textwidth]{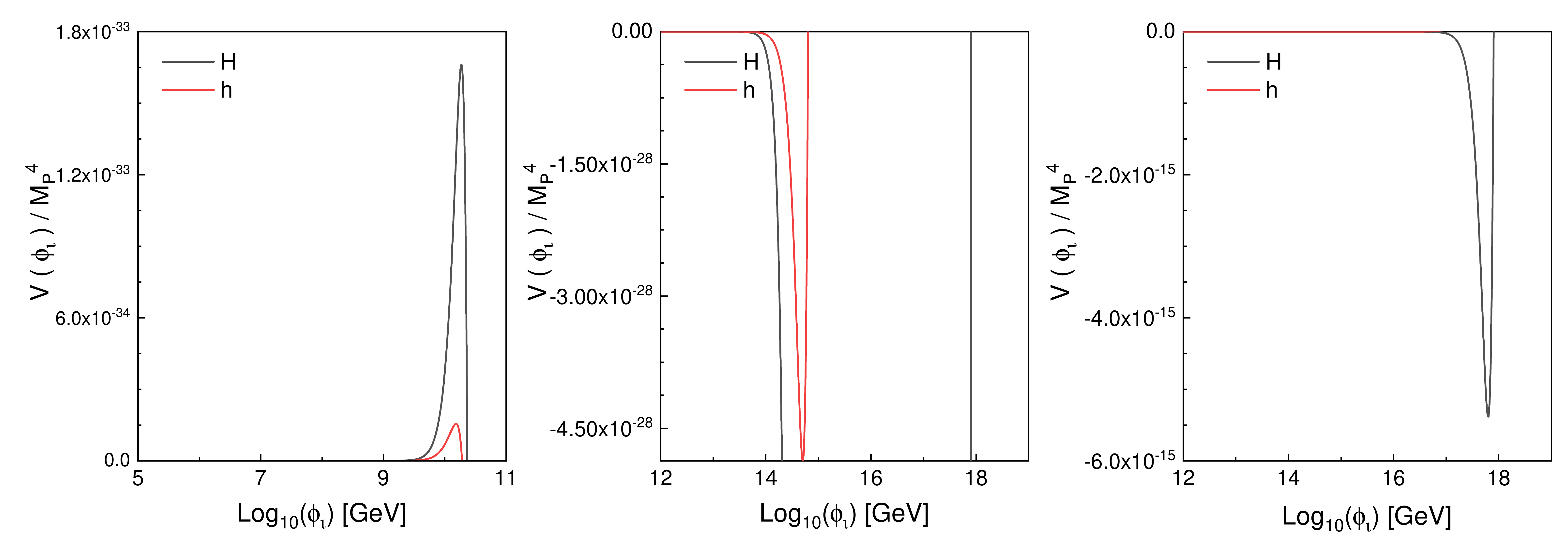}
\caption{\it The profile of the effective potential along the  $H$- and
  $h$-field directions, with
  $\tan\beta=50$ and $M_h^{\pm} = 500$ GeV. The potential along
  $H\, (h)$ is getting negative at scales around
  $\sim 10^{11}$\,GeV and reaches a new minimum at roughly
  $\sim 10^{18}$\,GeV ($\sim 10^{15}$\,GeV).}
 \label{vev-f}\end{figure}

Figure~\ref{vev-f} gives the profile of the MS-2HDM potential along
the $H$- (black curves) and $h$- (red curves) field directions.  By
considering the running quartic couplings, the instability shown in
Figure~\ref{vev-f} occurs when $\lambda_{2}$ crosses zero and becomes
negative after the first unification point~$\mu^{(1)}_X$. Note that the scalar potential along the field direction of the SM-like Higgs boson $H$ has a higher maximum and a deeper minimum compared to those found along the $h$ direction.

Table~\ref{tab3} presents the results of our analysis, which gives the
center of the bounce solution${}$~$\phi^{\rm b}_i(0)$ for the neutral CP-even
Higgs bosons $H$ and $h$, divided by the Plank mass
$M_P \approx 1.9 \times 10^{19}$\,GeV. The same table includes 
numerical estimates of the EW vacuum lifetimes $\tau_h$ and
$\tau_H$, as well as the values of the effective quartic couplings
$\lambda_{1,2,3,4}$ at the center of the bounce solution.

\begin{table*}[t]
\begin{center}
\begin{tabular}{ l c c c c c c c c c c c c c c c c }
 \hline \hline
&&&&$\phi_i(0)/M_P$ && \quad $\lambda_1$ \quad && \quad $\lambda_2$ \quad & & \quad $\lambda_3$ \quad && \quad $\lambda_4$ \quad &&\quad $\tau_i/T_U$\\ \hline
$H$ && &&$1.73 \times 10^{-5}$ & & $4.2 \times 10^{-4}$ && $-2.05 \times 10^{-3}$ & &$ -9.86 \times 10^{-3}$ && $2.818 \times 10^{-2}$&& $\sim 10^{5400\,\,}$\\ 
$h$ && && $5.06 \times 10^{-7}$ && $-1.6 \times 10^{-4}$& & $-1.11 \times 10^{-3}$ & & $-4.42 \times 10^{-3}$ && $1.162 \times 10^{-2}$&& $ \sim 10^{114180}$\\ 
 \hline \hline 
\end{tabular}
\end{center}
\caption{\it Numerical values for the center of the bounce solution
  divided by the Plank mass~$M_P$ for neutral CP-even Higgs bosons $h$
  and $H$, including their respective contributions to the EW vacuum
  lifetime~$\tau$ in units of the age of the Universe $T_U$. The values of
  the quartic couplings $\lambda_{1,2,3,4}$ at the center of the
  bounce solution are also given.}  
\label{tab3}
\end{table*}

We find that the EW vacuum lifetime $\tau$ is many orders of magnitude
larger than the age of the Universe, and even much larger than the SM
EW vacuum lifetime $\tau_{\rm{SM}} \sim 10^{640} \, T_U$ in the flat
space (in the absence of Plank-scale suppressed
operators~\cite{Branchina:2013jra,Branchina:2018xdh}). Therefore, we
safely conclude that our Universe is adequately stable and so our
high-scale quartic coupling unification scenario is phenomenologically
viable.

\section{Conclusions} \label{con}

We have considered one of the simplest realisations of a Type-II 2HDM,
the so-called Maximally Symmetric Two-Higgs Doublet Model (MS-2HDM).
The scalar potential of this model is determined by a single quartic
coupling and a single mass parameter. This minimal form of the
potential can be reinforced by an accidental SO(5) symmetry in the
bilinear field space, which is isomorphic to Sp(4)/$Z_2$ in the field
basis $\bf{\Phi}$ given in~\eqref{eq:bfPhi}. The MS-2HDM can naturally
realise the so-called SM alignment limit, in which all SM-like Higgs
boson couplings to $W^\pm$ and $Z$ bosons and to all fermions are
identical to their SM strength independently of $\tan\beta$ and the
mass of the charged Higgs boson.

The SO(5) symmetry of the MS-2HDM is broken explicitly by RG effects
due to the U(1)$_Y$ gauge coupling and equally sizeably by the Yukawa
couplings of the third generation of quarks and charged leptons.  For
phenomenological reasons, we have also added a soft SO(5)-breaking
bilinear mass parameter~$m^2_{12}$ to the scalar potential, which
lifts the masses of all pseudo-Goldstone bosons $h^\pm$, $h$ and $a$
above 500-GeV range in agreement with $B$-meson constraints. To
evaluate the RG running of the quartic couplings and the relevant SM
couplings, we have employed two-loop 2HDM RGEs from the unification
scale $\mu_X$ up to charged Higgs-boson mass~$M_{h^\pm}$. At the RG
scale~$M_{h^\pm}$, we have implemented matching conditions between the
MS-2HDM and the SM parameters.

Improving upon an earlier study~\cite{Dev:2014yca}, we have now
explicitly demonstrated that in the MS-2HDM all quartic couplings can
unify at much larger RG scales $\mu_X$, where $\mu_X$ lies between
$\mu^{(1)}_X \sim 10^{11}\,$GeV and $\mu^{(2)}_X \sim 10^{20}\,$GeV.
In particular, we have shown 
that quartic coupling unification can take place in two different
conformally invariant points, at which all quartic couplings vanish.
This property is unique for this model and can happen at different
threshold scales $M_{h^{\pm}}=500$\,GeV, $1$\,TeV, $10$\,TeV and
$100$\,TeV.  More precisely, the low-scale (high-scale) unification
point arises when $\lambda_2$ crosses zero and becomes negative
(positive) at the RG scale $\mu_X^{(1)}$
($\mu_X^{(2)}$).  The region between these two scales
corresponds to negative values of $\lambda_2$. For this reason, we
have performed a vacuum stability analysis of the model in order to
ensure that the EW vacuum is sufficiently long-lived. In this respect,
we have estimated the EW vacuum lifetime $\tau$ which was found to be
reassuringly long, i.e.~much larger than that of the EW vacuum
lifetime in the SM, $\tau_{\rm{SM}} \sim 10^{640} \, T_U$, assuming
the absence of harmful Planck-scale suppressed operators.

It is important to reiterate here that the MS-2HDM is a very
predictive extension of the SM, as it is governed by only three
additional parameters: (i)~the charged
Higgs-boson mass~$M_{h^{\pm}}$ (or $m^2_{12}$), (ii)~the ratio of VEVs $\tan\beta$,
and (iii)~the conformally-invariant quartic
coupling unification scale~$\mu_X$ which can only take two discrete
values: $\mu^{(1)}_X$ and $\mu^{(2)}_X$. Given these three
parameters, the entire Higgs sector of the model can be
determined.  In this context, we have presented illustrative predictions of
misalignment for the SM-like Higgs-boson couplings to the $W^\pm$ and
$Z$ bosons and, {\em for the first time}, to the top- and bottom-quarks.  The
predicted deviations to $Hb\bar{b}$-coupling is of order 10\% and may be
observable to future precision collider experiments.

Evidently, our novel theoretical framework can be straightforwardly
extended to multi-HDMs with $n$ scalar-doublets based on the
maximal symmetry group Sp($2n)/Z_2 \otimes \text{SU(2)}_L$. We plan to
report progress on this issue in an upcoming publication.

\section*{Acknowledgements}

\noindent
The work of AP and ND is supported in part by the
Lancaster–Manchester–Sheffield Consortium\- for Fundamental Physics,
under STFC research grant ST/P000800/1. 
\\
The work of ND is also supported in part by the Polish National Science Centre
HARMONIA grant under contract UMO- 2015/18/M/ST2/00518 (2016-2020).

\end{document}